# Advances in ultrafast plasmonics

*Alemayehu Nana Koya[1,2,¶], Marco Romanelli[3,¶], Joel Kuttruff[4,¶], Nils Henriksson[5,¶], Andrei Stefancu[6,¶], Gustavo Grinblat[7,¶], Aitor De Andres[5,¶], Fritz Schnur[5,¶], Mirko Vanzan[3,¶], Margherita Marsili[3,8,¶], Mahfujur Rahaman[9,¶], Alba Viejo Rodríguez[10], Tilaike Tapani[5], Haifeng Lin[5], Bereket Dalga Dana[11], Jingquan Lin[12], Grégory Barbillon[13], Remo Proietti Zaccaria[14], Daniele Brida[10], Deep Jariwala[9], László Veisz[5], Emiliano Cortes[6], Stefano Corni[3,15], Denis Garoli[14]\*,
and Nicolò Maccaferri[5,10]\**

[1]GPL Photonics Laboratory, State Key Laboratory of Applied Optics, Changchun Institute of Optics, Fine Mechanics and Physics, Chinese Academy of Sciences, Changchun, China
[2]Department of Physics, College of Natural and Computational Sciences, P. O. Box 138, Wolaita Sodo, Ethiopia
[3]Department of Chemical Sciences, University of Padova, via Marzolo 1, 35122 Padova, Italy
[4]Department of Physics, University of Konstanz, Universitaetsstrasse 10, 78464 Konstanz, Germany
[5]Department of Physics, Umeå University, Linnaeus väg 24, SE-90187 Umeå, Sweden
[6]Nanoinstitut München, Fakultät für Physik, Ludwig-Maximilians-Universität München, 80539 München, Germany
[7]Departamento de Física, FCEN, IFIBA-CONICET, Universidad de Buenos Aires, C1428EGA Buenos Aires, Argentina
[8]Department of Physics and Astronomy, University of Bologna, 40127 Bologna, Italy
[9]Department of Electrical and Systems Engineering, University of Pennsylvania, 19104 Philadelphia, Pennsylvania, United States
[10]Department of Physics and Materials Science, University of Luxembourg, 162a avenue de la Faïencerie, 1511, Luxembourg, Luxembourg
[11]Department of Physics, College of Natural and Computational Sciences, Jinka University, P. O. Box 165, Jinka, Ethiopia
[12]School of Science, Changchun University of Science and Technology, Changchun 130022, China
[13]EPF-Ecole d'Ingénieurs, 55 Avenue du Président Wilson, 94230 Cachan, France
[14]Istituto Italiano di Tecnologia, Via Morego 30, 16163 Genova, Italy
[15]CNR-NANO Istituto Nanoscience, Via Campi, 213/a, 41125 Modena, Italy
\*Corresponding authors: denis.garoli@iit.it; nicolo.maccaferri@umu.se
¶These authors contributed equally

**Abstract.** In the past twenty years, we have reached a broad understanding of many light-driven phenomena in nanoscale systems. The temporal dynamics of the excited states are instead quite challenging to explore, and, at the same time, crucial to study for understanding the origin of fundamental physical and chemical processes. In this review we examine the current state and prospects of ultrafast phenomena driven by plasmons both from a fundamental and applied point of view. This research area is referred to as ultrafast plasmonics and represents an outstanding playground to tailor and control fast optical and electronic processes at the nanoscale, such as ultrafast optical switching, single photon emission and strong coupling interactions to tailor photochemical reactions. Here, we provide an overview of the field, and describe the methodologies to monitor and control nanoscale phenomena with plasmons at ultrafast timescales in terms of both modeling and experimental characterization. Various directions are showcased, among others recent advances in ultrafast plasmon-driven chemistry and multi-functional plasmonics, in which charge, spin, and lattice degrees of freedom are exploited to provide active control of the optical and electronic properties of nanoscale materials. As the focus shifts to the development of practical devices, such as all-optical transistors, we also emphasize new materials and applications in ultrafast plasmonics and highlight recent development in the relativistic realm. The latter is a promising research field with potential applications in fusion research or particle and light sources providing properties such as attosecond duration.

# 1. Introduction: taming the nanoscale with ultrafast plasmonics

Harnessing light-matter interactions at both the nanometer (nm) and femtosecond (fs) scales is one of the most demanding challenges that many researchers belonging to different research fields, such as condensed matter physics, chemistry and materials science, are trying to resolve today. Moreover, the knowledge of both the spatial and temporal structure of an optical field interacting with a nanomaterial can enable a wider understanding of light-matter interactions at the nanoscale. This allows, for example, to realize energy-efficient and faster all-optical information processing with PHz bandwidth or develop novel toolboxes for tailored nanoscale photochemistry. The nanoscale exploration of ultrafast optical phenomena has been largely accelerated by employment of plasmonic nanostructures. Surface plasmons, both propagating and localized (SPPs and LSPs, respectively), are light-driven collective oscillations of free electrons at the interface between a conducting material and the dielectric environment. They are known for their unique feature to concentrate light into deep sub-wavelength volumes well-beyond the diffraction limit, thus providing unprecedented opportunities to control and manipulate light at the nanoscale.[1–11] In this context, recent advances in coherent light sources, time-resolved optical measurements, nanotechnology, materials science, and nanoengineering, as well as numerical simulations, have offered great opportunities to explore the interaction of ultrashort light pulses and plasmonic structures with the highest possible precision and resolution, giving rise to the emergence of *ultrafast plasmonics*.[12–16] This thriving research field aims at the development of compact optical devices that can generate, control, modulate, sense, and process ultrafast optical and electronic excitations at the nanoscale.[17–22] Further, the unique potentials of all dielectric photonic materials or 2D materials, such as transition metal dichalcogenides, has lately been recognized by the ultrafast plasmonics community.[23] As a result, in the past decade, the field of ultrafast plasmonics has become a rapidly growing field of study with plenty of opportunities in fundamental science and practical applications.

In general, ultrafast optical techniques are used to explore the physical and chemical phenomena below 1 picosecond (ps) using ultrashort light pulses, which are obtained with mode-locking techniques and can reach attosecond (as) time duration.[24] Such ultrashort pulses allow to measure in real-time the fastest, including sub-fs, processes taking place during chemical reactions as well as collective electron dynamics in atoms, molecules, and solids, allowing precise spectroscopy and nanoscale metrology. Specifically, the interaction of ultrafast pulses with solid-state nanostructures made of metallic materials or heavily doped semiconductors allows us to control the spatial and temporal evolution of the optical near field associated with the excitation of plasmons and their nonlinear optical properties.[25–35] In general, the excitation of a plasmon is one of the fastest processes in condensed matter physics. In fact, the timeframe of the plasmon excitation, usually defined by the inverse of the plasmonic resonance spectral width, is in the order of 100 as.[36] The plasmon relaxation time, the so-called plasmon dephasing process, is also ultrashort, in the 10 fs − 1 ps range, where the main physical process involved is the electron–electron (e-e) and electron-surface scattering (see **Fig. 1a**). Subsequently, energy is dissipated via electron–phonon (e-ph) relaxation (1 – 100 ps), and phonon–phonon (ph-ph) (100 ps – 10 ns) scattering.[37–39] In more detail, during the first 100 fs following Landau damping (exponential decrease as a function of time of charge waves in plasma or a similar environment),[40] the non-thermal distribution of electron–hole pairs decays either through re-emission of photons (radiative decay) or through carrier multiplication caused by e–e interactions (non-radiative decay). During this very short time interval $\tau_{nth}$, the hot carrier distribution is highly non-thermal. The hot carriers, which are electrons having energies larger than those of thermally excited electrons at room temperature, will redistribute their energy by e-e and e-ph scattering processes on a timescale $\tau_{el}$ ranging from 10 fs to 100 ps. Finally, the heat is transferred to the surroundings of the metallic structure on a longer timescale $\tau_{ph}$ ranging from 100 ps to 10 ns, via thermal conduction.[41] Such ultrafast dynamics make plasmonic nanostructures promising for various applications, especially for enhanced spectroscopy, photocatalysis, information processing, and quantum technologies.

These and many other applications require precise spatial and temporal control over the optical responses of nanostructured materials on both the fs and nm scales. To meet this requirement, and to have a detailed insight into the physical mechanisms involved, a number of theoretical models,[42,43] in particular finite element method (FEM), finite-difference time domain (FDTD) method and time-dependent density functional theory (TDDFT), have been developed.[44–49] As well, experimental techniques, such as time-resolved photo-emission electron microscopy (TR-PEEM) or photo-induced near-field electron microscopy (PINEM), both fulfilling spatial and temporal resolution requirements[50–57], have been widely employed to investigate, and even shape, the ultrafast optical near-field dynamics of plasmonic nanostructures.[58–60] From an experimental point of view, it is also common to employ [1–6]optical pump-probe schemes, including X-ray free electron lasers (XFELs), to study the ultrafast charge and spin dynamics in plasmonic nanostructures, in particular those displaying magnetic properties.[61–66] Full-field-resolved detection of the temporal response of plasmonic nanostructures, as well as their nonlinear optical properties, in the mid infrared regime has also been carried out recently by employing time-resolved spectroscopy implementing electro-optical sampling (EOS) detection scheme.[67] Together, recent efforts in active manipulation of the optical response of nanostructured materials, rational design of the nanostructures and nanoscale engineering of their environment, as well as material composition, offer an exciting opportunity to develop ultrafast, all-optically reconfigurable nanophotonic devices.[68–73]

As a result of the aforementioned developments in the field, this review showcases the most recent and important theoretical and experimental advances in ultrafast plasmonics, and it is structured as follows. **Section 1** gives a general introduction to the topic, followed by **Section 2**, where we provide a broad and updated overview of modelling and experimental characterization techniques used in the field of ultrafast plasmonics. **Section 3** focuses on the challenges and opportunities of ultrafast plasmon-driven chemistry with particular focus on plasmon-enhanced photocatalysis. **Section 4** highlights the most recent advances in ultrafast multi-functional plasmonics, providing insights on how to leverage on charge, spin, and lattice degrees of freedom for active and ultrafast control of optical and electronic properties at the nanoscale, by utilizing either plasmon or phonon polaritons. **Section 5** provides insights into new materials and the most common applications of ultrafast plasmonics.

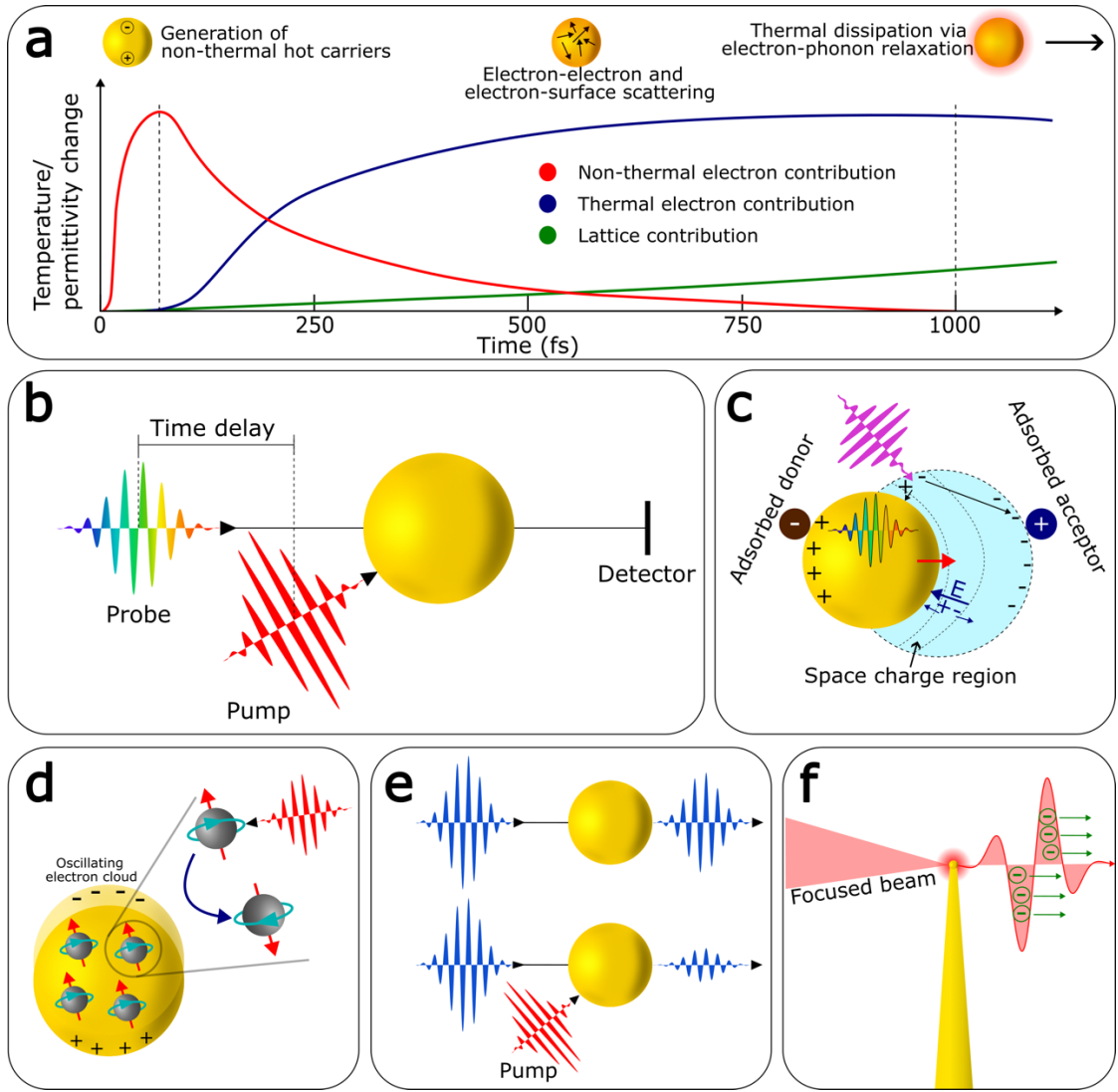

**Figure 1. Conceptual illustrations of the different sections of the review.** (a) Illustration of the timescales involving carrier relaxation processes following LSP excitation (the same concept can be applied also to SPPs). First, non-thermal hot carriers are generated (1-100 fs). Afterwards, their energy dissipates into thermal energy via electron-electron, electron-surface and electron-phonon scattering, causing a rise in electron and lattice temperatures. The change in electronic temperature reflects in a change of the permittivity of the structure. (b) In Section 2 we give an overview of the most recent developments in modelling and experimental characterization of ultrafast plasmon dynamics, for instance by performing optical pump-probe experiments. (c) Section 3 is dedicated to plasmon-driven chemistry. One of the main applications discussed is plasmon-enhanced photocatalysis. (d) In Section 4 we introduce the reader to active ultrafast plasmonics, where materials possessing different properties (for instance both plasmonic and magnetic, the so-called magnetoplasmonic materials) are used to control spin dynamics at the nanoscale. (e) Section 5 is dedicated to the most common applications of ultrafast plasmonics, in particular plasmon-driven switching of optical signals. (f) Section 6 focuses on the rising field of relativistic plasmonics, where high intensity ultrashort light pulses can excite strong plasma oscillations in metallic nano-tips resulting in matter ionization, and a consequent coherent light-driven acceleration of electrons.

Finally, **Section 6** gives a brief overview of the raising field of relativistic plasmonics, where laser intensity reaches $10^{11}$ - $10^{13}$ W/cm$^2$. For sake of clarity, a schematic overview of the review main contents is shown in **Fig. 1b-f**. We conclude by giving our personal vision on the prospects of the several research areas in ultrafast plasmonics covered in this review. Despite our best attempt to report accurately the most important advances in the field and do justice to all its novel developments and its diversity, the research area is expanding so fast that there remains great latitude in deciding what to include in this review, which in turn means that some areas might not be adequately represented here. However, we feel that the sections that form this review, each written by experts in the field and addressing a specific subject, provide an accurate snapshot of where this research field stands today. Correspondingly, it should act as a valuable reference point and guideline for emerging research directions in ultrafast plasmonics, as well as illustrate the directions this research field might take in the foreseeable future.

## 2. Experimental characterization and modelling of ultrafast plasmon dynamics

Metallic nanostructures featuring both SPPs and LSPs, are highly promising because of their ultrafast dynamics and potential to confine electromagnetic (EM) field in sub-nanometric spatial regions.[74–77] Understanding and controlling plasmons is crucial for advancements in a wide range of applications including sensing, spectroscopy, photocatalysis, nanoscale metrology, ultrafast computation, and renewable energy.[78–84] However, the different length and time scales involved make the study of these effects rather complex. To fully understand the physics underlying ultrafast plasmonic processes, multiple experimental and modelling approaches have been developed over the last decade, each one providing key insights into the general picture. In this section, we provide a concise yet comprehensive overview of these experimental and theoretical methodologies.

The ability to image plasmon near-fields, that is EM fields between 1 and 10 nm from the plasmonic structures surface oscillating with the period of few light cycles, is crucial to guide future directions in ultrafast plasmonics research. Such techniques require both sub-cycle time and sub-wavelength spatial resolution, spanning the so-called nano-femto scale. One widely used approach is TR-PEEM , which has been successfully applied to image both SPPs and LSPRs.[85–87] In such experiments, a short light pulse (called pump) excites plasmons in a nanostructured material. A time-delayed phase-stable duplicate of the first pulse then interferes with the plasmons and promotes photoelectrons to the vacuum through a multi-photon photo-emission process. Electrons are subsequently collected using electron objective/projective and an electron sensitive camera in a low-energy electron microscope, providing the necessary nm spatial resolution. In the past, this technique has been applied to image the dynamics of both LSPs and SPPs in metallic nanostructures.[88–94], providing also a way to understand more deeply the most important factors in plasmon damping.[95] Furthermore, TR-PEEM has been used to study nanofocusing of SPPs,[96] with many potential applications including heat-assisted magnetic recording,[97] or plasmon-induced electron emission.[98,99] Recently, TR-PEEM was advanced by Davis *et al.* to include full vectorial mapping of the plasmonic near-field,[100] as shown in **Fig. 2a**. In their experiments, a light pulse excites SPPs from grooves in the sample that lead to non-trivial plasmonic fields. The plasmons then interfere with the probe pulse that arrives on the sample with an adjustable time delay and frees photoelectrons into vacuum. In contrast to previous approaches, they perform a set of two independent measurements with orthogonal polarization of the probe pulses. This allows to retrieve the full vectorial composition of the plasmonic field, making possible the demonstration of plasmonic spin-momentum locking, as well as excitation of plasmonic skyrmions. The proposed vector microscopy promises the exploration of exciting phenomena associated with spin-photon coupling and orbital angular momentum physics. Noteworthy, PEEM can be used not only for imaging plasmon dynamics, but also to study charge carrier dynamics on semiconductor surfaces.[101–104] Hence, we foresee that TR-PEEM will be instrumental to unravel plasmon-generated hot-carrier dynamics on surfaces of novel plasmonic media (e.g., heavily doped semiconductors) beyond the traditional noble metals.

Another efficient approach for plasmon imaging is the ultrafast transmission electron microscopy (U-TEM).[105–108] In this case, the sample is excited by a light pulse from a fs-laser source, and probed by a time-delayed electron pulse generated by photo-emission inside an electron microscope using the same laser source. While using TEM nm spatial resolution can be achieved, creating short electron pulses and keeping them from spreading in time before reaching the sample is very challenging due to their intrinsic dispersion in vacuum.[109] For this reason, traditionally radio-frequency cavities are employed for the compression of electron pulses, but the need for active synchronization leads to unavoidable timing-jitter of the compressed pulses.[110–113] A way to circumvent active synchronization is to use the cycles of laser-generated THz-fields for electron compression in an all-optical scheme.[114–117] Such THz pulses can be generated from the same laser source that is employed for photo-emission, giving possibility to suppress timing jitter to sub-fs timescale.[118] U-TEM has been successfully used to study acousto-optic dynamics in plasmonic nanoparticles,[119] and polaritonic strain waves in $MoS_2$ nanoflakes propagating at approximately the speed of sound (7 nm/ps).[120] Nevertheless, to date electron pulses inside a U-TEM are still around two orders of magnitude too long to measure plasmon charge oscillations directly. However, when interacting with the sample, the electrons inside an electron pulse can absorb a photoexcited plasmon or excite a second plasmon through stimulated electron energy loss in a quantum-coherent fashion, i.e., considering the wave nature of the electron. Hence, via post selection of only the transmitted electrons that have gained/lost energy, the plasmonic near-field can be imaged using the PINEM technique. Following the pioneering work of Barwick *et al.*,[50] PINEM has been used for near-field imaging of many different nanostructures, including metallic nanoparticles[121,122] and nanoantennas.[123–125] Furthermore, the coherent interaction with the light field in PINEM was used to study and prepare tailored quantum states of free-electrons.[57,126–134] Although sub-optical-cycle resolution is not readily available in current PINEM setups, which prevents at the moment the measurement of plasmonic fields directly in space and time, plasmon interference in PINEM can be used to reveal properties of plasmon propagation and decay.[135] Recently, this direction was followed by Madan *et al.*, who used holographic plasmon imaging to quantitatively measure phase and group velocity of SSPs with as and nm resolution (see **Fig. 2b**).[54] In contrast to conventional PINEM, where only one optical pump pulse is used, they employed two pump pulses, of which each launches a SPP (see upper panel in **Fig. 2b**). By varying the time delay between the pump pulses, they could control the plasmon interference pattern, allowing to measure the corresponding SPP propagation parameters (see lower panel in **Fig. 2b**). Finally, the perspective to sample plasmonic fields using PINEM with as electron pulse trains promises even higher temporal resolution and superior sensitivity, making such approach promising for future studies of photonic nanostructures at ultimate dimensions in space and time.[136–138]

Beyond the possibility to generate coherent charge oscillations, plasmon-generated hot charge carriers represent a rich source of localized energy that can be harvested, for instance, either in photovoltaic devices or supplied to adsorbed molecules to drive chemical reactions. This change of electronic band-population induced by plasmon decay leads to transient changes in the optical properties of a material that can be tracked using short light pulses. To this end, optical pump-probe spectroscopy is ideally suited to study the redistribution of electronic energy on sub-ps time scales. In such experiments, a light pulse excites a sample (pump), and a second, time-delayed pulse is used to measure changes in the optical response of the system (probe). Due to stroboscopic sampling, the time-resolution in pump-probe experiment is not limited by the response time of the detector, but rather determined by the duration of the optical pulses that are used.[139] The need to resonantly excite plasmons at different frequencies and detect broad spectral signatures of transient states calls not only for the shortest durations but also for broad tunability of pump and probe pulses. In this context, the development and commercial availability of novel laser sources and, in particular, next-generation optical parametric amplifiers (OPAs) has greatly accelerated research in the field in the last few decades, as they can provide ultra-tunable pulses with few-fs time durations.[140,141] Spectroscopy systems based on Ti:Sapphire technology are well established nowadays in many labs, as short pulses below 10 fs can be directly obtained from the laser oscillator.[142] Furthermore, Ti:Sapphire-driven OPA systems have been demonstrated with carrier frequencies ranging from ultraviolet to mid- and far-infrared.[143–145] More recent efforts also focused on developing alternative systems based on ytterbium laser sources, as they can provide excellent stability and tunable repetition rate up to several hundreds of kHz.[146]

In a pioneering study by Sun *et al.*, pump-probe experiments were performed to study the dynamics of hot electrons in thin gold films.[147] Since then, many works followed focusing on the nonlinear optical response of metals[148] and, in particular, plasmonic nanostructures.[130–137] Della Valle *et al.* have used pump-probe spectroscopy to investigate plasmon detuning and other physical mechanisms dominating the nonlinear plasmon dynamics in individual plasmonic antennas (**Fig. 2c**),[149] and more recently they have introduced novel concepts for ultrafast active control of the optical response of plasmonic metasurfaces[47,73] and metagratings.[71] In this context, it is worth mentioning the pioneering work by Taghinejad *et al.*, who demonstrated the active ultrafast modulation of the phase, polarization, and amplitude of light through the nonlinear modification of the optical response of a plasmonic crystal that supports subradiant, high Q, and polarization-selective resonance modes (**Fig. 2d**).[150] Noteworthy, the time-resolution of the pump-probe experiments can be pushed even into the sub-fs timescale by employing high harmonic generation.[151–154] Combining XUV attosecond pulses with NIR pumping, Niedermayr *et al.*

recently studied nonequilibrium optical properties of aluminum, revealing few-fs dynamics of hot charge carriers relevant for ultrafast plasmonics applications in the UV.[155] By leveraging on nm-gap size in bow-tie plasmonic nanocircuits, Ludwig et al. have shown sub-fs resolution by tuning the interference of carrier-envelope-phase-stable single-cycle infrared pulses (see **Fig. 2e**).[46,156] Interferometric correlation measurements showed control of the transfer of individual electrons between two metallic nanocontacts, which can be further controlled by applying an external dc bias over the gap.[157] It is worth to highlight here that all these studies open new avenues of investigating plasmonic integrated surfaces for optical logics operating at PHz speed frequencies. In parallel with the development of advanced technologies to implement complex time-resolved experiments to access the ultrafast dynamics in plasmonic systems, theory has been instrumental in corroborating experimental findings and suggesting novel plasmon-related phenomena that were subsequently probed experimentally. Most of the models presented so far can be mainly grouped into three broad categories: (i) those studies that employ a classical continuum-based description of the nanoparticle (NP), whose optical response is dictated by its shape and dielectric function; (ii) those that describe the NP using jellium model, where the electronic wavefunction explicitly include only valence electrons, while the nuclei and the core electrons are considered as a homogeneous frozen background; and (iii) those that include explicitly the nuclear and electronic structure using ab-initio techniques. Classical electrodynamics-based methods, such as FDTD[158] and Time-Domain Boundary-Element-Method (TD-BEM)[47,159–161] have been used to investigate the real-time dynamics of plasmonic materials (see **Fig. 3a,b**), even their interaction with nearby molecules.[162,163] Building on such descriptions, classical models based on modified bulk dielectric functions in combination with kinetics rate equations[164–166] have been extensively employed to interpret pump-induced ultrafast nonlinear responses in plasmonic systems.[149,167] In particular, the three-temperature model (3TM) was used.[149,148,168] This rate equation model describes the energetic relaxation process of the nanostructure in terms of three degrees of freedom, i.e., the density of excess energy stored in the hot carriers population created by plasmon decay, the thermalized electron gas temperature and the lattice temperature.[169,170] This approach has been successfully applied to simulate the ultrafast transient absorption spectra of gold-nanoparticles,[161] making it possible to rationalize the findings of state-of-the-art experimental data. Moreover, following the preliminary approaches based on Lorentz oscillator model[42] and time-domain resonant-mode-expansion theory,[43] Lin et al. have developed various active control approaches including coherent control techniques,[44,49] polarization manipulation,[55] and adaptive laser pulse shaping methods[48] to control the ultrafast dynamics in complex plasmonic nanostructure geometries.

A seminal work based on a jellium model by Manjavacas et al., aiming at describing the plasmon-induced hot-carriers generation rate dates back to 2014.[171] Since then, other methodologies based on a jellium-like description of the NP tried to develop a time-resolved picture of the electronic dynamics upon plasmon excitation, including e.g. e-e and e-ph scattering effects (see **Fig. 3c**).[172–179] Despite the undoubted usefulness of jellium models, they cannot account for the real material atomistic structure (and the resulting band structure), that has been shown to be relevant for the quantitative estimation of transport properties, lifetimes, as well as energy and momentum distributions of hot carriers.[169,180–182]

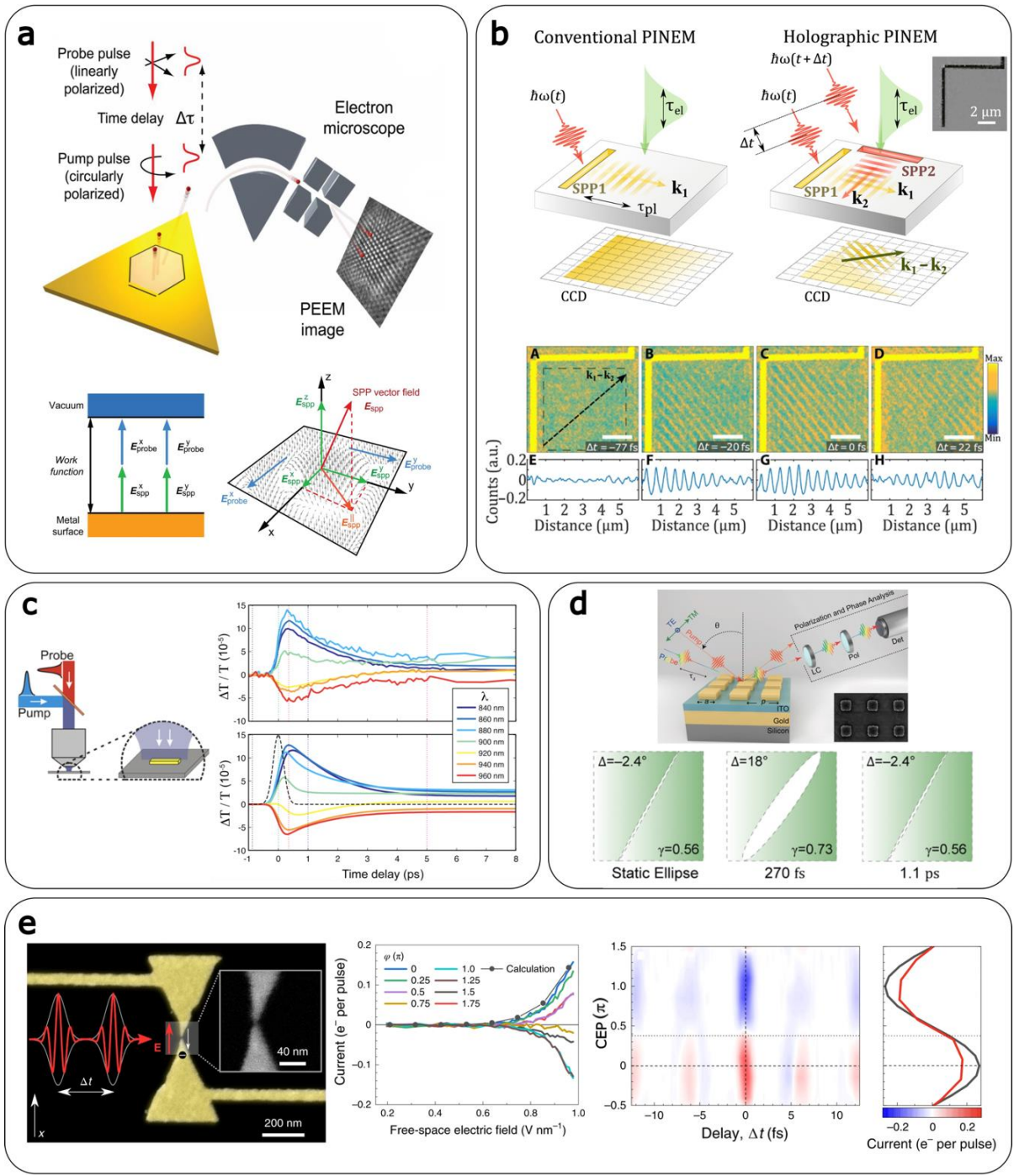

**Figure 2. Different methodologies to study and control ultrafast optical and electronic dynamics in plasmonic architectures.** (a) Ultrafast time-resolved vector microscopy of plasmonic fields using photo-emission electron microscopy (PEEM). Two-photon-PEEM process is used to obtain vector and time information from surface plasmons. Vectorial components are obtained via the two orthogonal probe fields during two separate measurements, which allows imaging of plasmonic skyrmions in three dimensions.[100] Reprinted with permission from AAAS. (b)

Holographic photon-induced near-field electron microscopy (PINEM). In contrast to conventional PINEM that images the time-averaged propagating SP envelope, two propagating SPs form a standing wave that can be imaged as a periodic modulation of PINEM intensity. Holographic images obtained at different delays between the excited SPs are used to determine their phase and group velocities.[54] Reprinted with permission from AAAS. (c) Pump-probe spectroscopy employed to study the dynamics of single gold nanoantennas. Experimental differential transmission (upper) of the plasmonic dynamics when pumped by a 780 nm wavelength laser is approximated using numerical simulations based on the 3TM (lower). The colors represent different probe wavelengths.[149] Reprinted with permission from ACS. (d) Ultrafast control of phase and polarization. Schematic of a plasmonic crystal consisting of gold nanoparticles on a layer of indium tin oxide, placed on a gold film. The pump-probe spectroscopy setup is used to estimate the phase and polarization of the crystal, estimated from the reflected light using a liquid crystal phase retarded (LC) and a polarizer (POL), respectively. The polarization dynamics (lower) are measured at different time delays when the crystal is excited by an on-resonance pump.[150] Reprinted with permission from ACS. (e) Ultrafast currents in a nanocircuit controlled by single-cycle light pulses. The carrier-envelope phase (CEP) of the biasing pulses can be tuned to control the transfer of individual electrons in the nanocircuit. Interferometric current autocorrelations via two driving pulses allow to study electron transport at sub-fs timescales.[46] Reprinted with permission from Springer-Nature.

For this reason, there has been growing interest in applying ab-initio methods to nanometallic systems, despite their higher computational cost. Most of these studies rely on approaches such as DFT and others like the embedded correlated wavefunction (ECW) method.[183,184] In this case the systems is split into two regions, where the most interesting one is treated through high level quantum mechanical theories in presence of an embedding potential obtained by a DFT calculation, mimicking the presence of the surroundings. Different studies have described the time-resolved electron dynamics by employing real-time Time Dependent Density Functional Theory (rt-TDDFT), where the response of the system is simulated by integrating in time the time-dependent Kohn-Sham equations in the presence of an external pulse (see **Fig. 3d**).[185–192] Many Body Perturbation Theory (MBPT) approaches used to calculate the self-energy of a many-body system of electrons, such as the GW approximation, have also been used to describe the e-e scattering contribution to the carrier lifetimes,[193] also in combination with non-linear Boltzmann equations to model the time-resolved dynamics.[194] Along these lines, more recently, first-principle real-time non equilibrium Green's function formulations fully including e-e and e-ph scattering have been developed,[195] laying the foundation for upcoming promising works in the field.

### 3. Ultrafast plasmon-driven chemistry

Recently, our society is striving to do, in a few decades, what nature has achieved in millions of years in order to address some of humanity's greatest challenges, for instance decreasing the greenhouse gas emissions and, at the same time, generating 'green' energy. To this end, designing photocatalysts based on plasmonic structures is a promising method to achieve efficient light-to-chemical energy conversion in the visible range of the electromagnetic spectrum.[83,196,197] As discussed in **Section 1**, once the plasmon resonances have been excited, the energy of the oscillating electron plasma decays either radiatively, by re-emitting light (with a rate proportional to the nanoparticle's volume) or non-radiatively, by generating energetic electron-hole pairs (hot carriers).[3] Then the temporal evolution of the hot carriers' distribution follows the well-known timeline, described by the three-temperature model (see **Section 2**),[147,198] which sets a time limit after which we can no longer extract the energy of the hot carriers, e.g., driving a chemical reaction, since that energy is lost as heat.

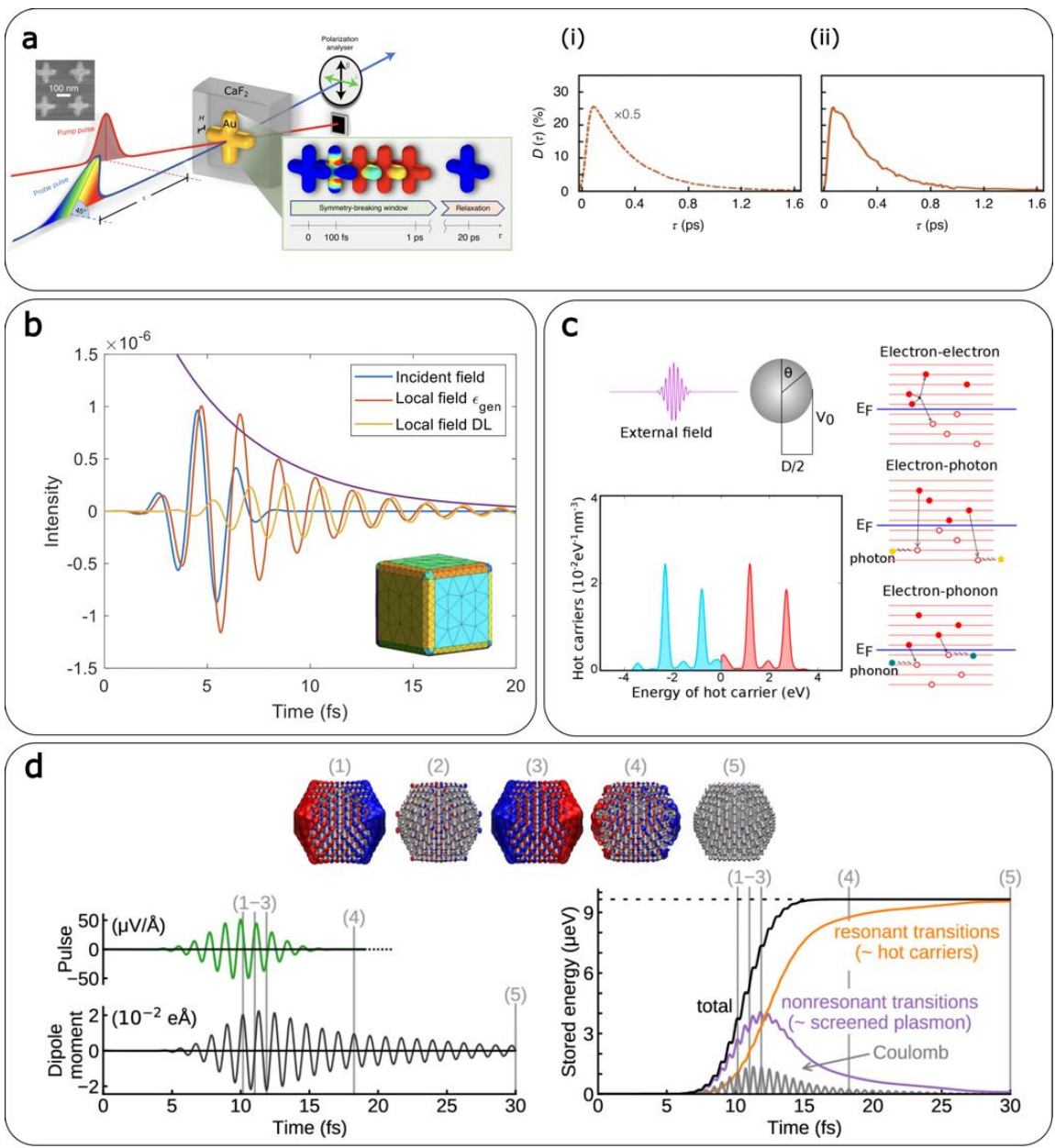

**Figure 3. Theoretical approaches aimed at investigating plasmon dynamics and plasmon-induced effects.** (a) Sub-ps broadband dichroism recovery achieved utilizing spatial inhomogeneities of photogenerated hot carriers in a symmetric gold nanostructure. Schematic of the pump-probe spectroscopy used to measure the broadband dichroism dynamics of the structure. The inset is an illustration of the inhomogeneous change in imaginary permittivity at different time delays. The dichroic ratio is (i) estimated using a numerical model based on the FEM and a spatially dependent 3TM and (ii) determined from experimental dynamic measurements of polarization of the structure.[47] Reprinted with permission from Springer-Nature. (b) Time evolution of the incident and local fields nearby a gold nanocube featuring a side length of 10 nm whose time-resolved response is obtained by TD-BEM. The plasmon-

induced local field is computed assuming two different dielectric function models, namely, Drude-Lorentz-like (yellow curve) and the generic dielectric function approach (orange curve). The excitation pulse has a frequency of 2.07 eV and it is represented by a sinusoidal wave modulated by a gaussian envelope.[160] Reprinted with permission from AIP. (c) Plasmon-induced hot carriers generation in a silver jellium NP having a diameter of 6 nm under resonant excitation at 3.5 eV. The lower-left panel displays the hot electrons (red) and hot-holes(blue) distribution per unit volume as a function of energy relative to the Fermi level. The relaxation channels for the hot carriers that were considered are shown on the right: electron-electron, electron-photon (radiative recombination) and electron-phonon scattering.[173] Reprinted with permission from ACS. (d) Real-time dynamics of the plasmonic response of a silver Ag561 NP by means of rt-TDDFT. The incoming pulse profile is shown on the left (green curve) along with the NP dipolar time-dependent response (black curve). Electron density oscillations are shown on top at different time intervals (1-5) where the red and blue colors stand for density increase and decrease respectively. The lower right plot shows the time evolution of the total energy stored in the excited system and its decomposition in terms of non-resonant transitions ("screened plasmon contribution", purple curve), resonant transitions ("hot carriers contribution", orange curve) and Coulomb energy (grey).[190] Reprinted with permission from ACS.

As it became increasingly clear that metal plasmonic nanostructures would always be associated to high energy losses,[199] hybrid metal-semiconductor structures have been proposed as great candidates to circumvent this issue.[200,201] The idea is that, instead of using directly the plasmonic nanoparticle to drive a chemical reaction, it could be used to harness the visible light, generating hot carriers, which can be transferred to nearby semi-conductor (such as $TiO_2$), that would otherwise not absorb light in the visible range. To export hot electrons to a nearby semiconductor for any practical application, the charge transfer process needs to be completed at a rate faster than the electron-phonon scattering process. To facilitate hot electron injection, three fundamental aspects need to be fulfilled: i) sufficient initial hot electrons energy to overcome the interfacial Schottky barrier, ii) direct contact between metal and the semiconductor, and iii) minimal lattice defects and impurities at the interface. The semiconductor then would drive the chemical reaction. There are two mechanisms for charge transfer from the metal nanoparticle to acceptor states (either belonging to a molecule or to a semi-conductor): a direct charge transfer upon excitation by light (which is also at the basis of the so-called chemical interface damping of the plasmon) and a sequential mechanism where plasmons first decay into hot carriers through surface-assisted Landau damping, and then tunnel to the acceptor states.[202,203] Because of the Schottky barrier formed at the metal nanoparticle/semiconductor interface, the hot carriers are not likely to be transferred back to the metal.[204,205] In this context, ultrafast optical spectroscopy allows us to monitor directly the dynamics of hot carriers at the metal-semiconductor interface,[205] and to design more efficient plasmonic photocatalysts.[206] A seminal work in the field of hybrid plasmonic nanostructures is the study of Wu *et. al.*,[207] who designed hybrid Au-cadmium selenide (CdSe) nanostructures with a quantum yield of 24% for light energy over 1 eV. They highlighted the presence of a direct charge transfer pathway from Au to the conduction band of CdSe, by which the plasmon resonance in gold decays directly into electron-hole pairs, with the electron localized in the conduction band of CdSe. By performing ultrafast measurements, they were able to monitor the dynamics of plasmon-induced hot carriers in the hybrid Au-CdSe nanostructures. The charge transfer to the CdSe takes place in 20 fs, consistent with the direct charge transfer mechanism, while the back charge transfer, from CdSe to Au, took place in about 1.45 ps. This seminal work showed that a high charge transfer quantum yield can be achieved through a direct charge transfer pathway, avoiding the usual losses associated with plasmonic chemistry.[207,208] Theoretical modelling is also pivotal in characterizing the charge transfer mechanism in some cases. Zhang *et al*. studied Au nanorods close to $MoS_2$, by combining rt-TDDFT with FSSH for non-adiabatic nuclear effects.[209] They were able to show the presence of an indirect charge transfer process in which a fast plasmon decay (< 30 fs) first led to hot carriers formation in the NP, followed by their injection into $MoS_2$ within 100 fs (see **Fig. 4**). Direct and indirect mechanisms can also coexist as shown by Long *et al*. for Au nanoclusters close to a $TiO_2$ slab who also suggested that the degree of coupling between the NP and the semiconductor could strongly influence the type of mechanism observed.[210]

As opposed to the very fast dynamics of charge carriers in plasmonic (hybrid) catalysts, that take place in the fs timescale, the motion of nuclei in molecules happens on the ps timescale while the kinetics of chemical reactions takes place on the µs to sec timescale (nine to fifteen orders of magnitude slower than hot carriers thermalization).[206] The difference between the characteristic timescales of hot carriers and chemical reaction dynamics is perhaps the most important bottleneck in plasmonic chemistry.[83,206] This rough timescale analysis illustrates why ultrafast optics is mostly used to probe the dynamics of charge carriers in (hybrid) catalysts, and not in molecules: chemical reactions simply do not proceed on such short timescales. Of course, this does not mean that there are no processes taking place on the fs time scale (or even faster) in molecules.[211] Starting with the well-known work of A. Zewail in fs chemistry,[212] this field has been a prospering one, investigating fundamental ultrafast processes in molecules. Indeed, over the last few years there has been growing interest in affecting chemical reactions at short time scales,[213] for instance by means of strong light-matter coupling, where the formation of hybrid light-matter polaritonic states leads to a sub-ps rearrangement of the molecular electronic energy levels that can in turn affect the subsequent photochemistry and reaction dynamics.[214–220] To give an example, optical cavities have been used to suppress or enhance chemical reactions both in ground and excited states, resulting in interesting chemical applications such as singlet fission[221] and selective isomerization.[222–224] Despite the fact that strong light-matter coupling can also be achieved using plasmonic systems[223], the field is still in its infancy, in particular on the experimental side, especially if compared to the use of polaritonic optical cavities, for which ultrafast investigations already appeared.[225] For these reasons, in what follows we omit a detailed discussion of ultrafast investigation of chemical reactions modified by plasmon-molecule strong interactions and we will focus on plasmonic chemistry related to chemical reactions that are usually much slower than the electronic processes taking place in heterogeneous catalysts (such as molecular dissociation, desorption, diffusion).[206] Thus, in the following we focus on the applications of ultrafast optical processes in plasmonic (hybrid) catalysts, highlighting how ultrafast optics can direct the design and understanding of the fundamental processes taking place in these materials.[196,204]

In one exemplary study on Au/Pt core-shell NPs, Engelbrekt *et. al*. tracked the transfer of energy from the Au core to the Pt shell, where chemical reactions can take place.[226] By carefully monitoring the Pt shell thickness through photoelectron spectroscopy, the authors were able to finely control the content of Pt, from sub-monolayer to multiple layers. The most important finding of this study is that the energy of the plasmon generated hot carriers can be transferred on the sub-ps scale to the Pt shell, where it can drive chemical reactions. Although the addition of the Pt shell leads to a faster dephasing of the plasmon resonance due to the collision of the electrons participating in the plasmon resonance with the Pt electrons and faster charge transfer, the electron density in the Au core and the LSPR oscillator strength remain fairly constant. This is important since it confirms that indeed the plasmon resonance is only due to the electrons in the Au core, without contributions from the Pt electrons. Moreover, the broadening of the AuNPs LSPR due to the addition of Pt should not negatively influence the photo-catalytic efficiency of Au/Pt NPs with broad-band excitation (such as solar light). Engelbrekt *et. al.* used both spectral and time domain data to get information on the ultrafast response of the Au and Au/Pt core-shell NPs. In the spectral domain, the authors were able to extract information about the earliest time-dynamics, the plasmon dephasing time (see **Fig. 5a**).[226] As expected, with increasing Pt content, the plasmon dephasing rate increased due to interfacial damping. After the plasmon dephasing on the fs scale, the dynamics of hot carriers could be tracked directly in the time domain. The hot carriers' dynamics is characterized by a faster decay time, due to electron-phonon coupling (thermalization), and a slower decay time of the phonon-phonon coupling, leading to the cooling of Au and Au/Pt NPs.

Another study highlighting the use of ultrafast optics to design more efficient plasmonic photocatalysts was reported by Kumar *et. al.*[227] The authors showed that gold nanoparticles (AuNPs) covered with reduced graphene oxide (r-GO) are more effective than platinum (Pt)-covered AuNPs at converting $CO_2$ to formic acid (HCOOH). With a selectivity toward HCOOH > 90%, the quantum yield of HCOOH using r-GO-AuNPs is 1.52%, superior to that of Pt-coated AuNPs (quantum yield: 1.14%), when excited close to the surface plasmon resonance of the AuNPs, at 520 nm. The reason for the higher quantum yield of the r-GO-AuNPs is the higher mobility of electrons in the graphene monolayer, which favors a higher rate of electron transfer to adsorbed $CO_2$. By using $TiO_2$ as an electron acceptor, they found an electron transfer time of less than 220 fs for the r-GO-AuNPs, leading to a charge transfer rate 8 times higher than for AuNPs, and 4 times higher than GO-AuNPs (see **Fig. 5b**). Therefore, by optimizing the ultrafast response of hybrid Au-GO plasmonic nanostructures, the design of cheaper and more efficient photocatalysts could be achieved. Interestingly, Hoggard *et. al.* also analyzed the ultrafast charge transfer from AuNPs to graphene monolayers. By monitoring the plasmon resonance linewidth of single AuNPs deposited on graphene, they were able to extract a time of 160 fs for the direct metal-graphene charge transfer, and estimate that such a transfer has an efficiency of 10%.[228] In a similar study, Sim *et. al.*[229] studied the ultrafast dynamics of plasmon generated hot carriers in AuNPs combined with more traditional catalytic materials, such as Pt and Ni. An important feature that was taken into consideration is the capping agent or ligands on the metal nanoparticles. It was found that the adsorbed molecules can influence the ultrafast dynamics of the plasmon generated hot carriers through, for example, direct metal-molecule charge transfer.[230,231] In particular, in this study by Sim *et. al.*, the authors synthesized the bi-metallic Au/Pt and Au/Ni nanoparticles by annealing a 5 nm thick Au thin film fabricated by e-beam evaporation, after which either Pt or Ni thin films, were deposited. By looking at the different pump-probe measurements of the differential transmission spectra, it was found that both Au/Pt and Au/Ni NPs have significantly faster decay dynamics than bare AuNPs. The differential transmission spectra for the three different samples were fitted with a two-temperature model (i.e., two decay components).

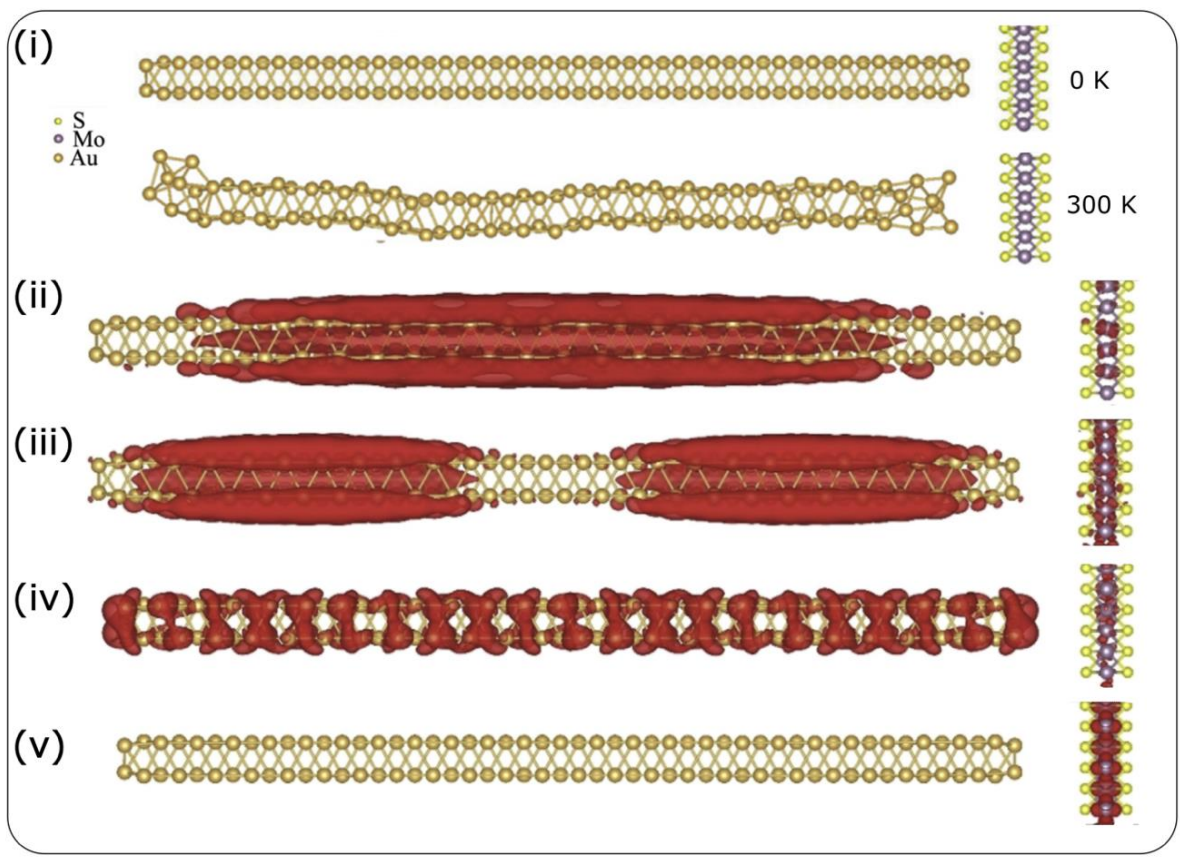

**Figure 4. Computational investigation of hot-electron injection from Au nanorods into MoS₂ combining rt-TDDFT and nonadiabatic molecular dynamics.** (i) Optimized structure of the $Au_{100}$-$(MoS_2)_{12}$ system under study at 0 K, and one representative geometry obtained from the molecular dynamics simulation run at 300 K. Thermal fluctuations leads on average to an increase of the metal-semiconductor distance, thus reducing the strength of their interaction. (ii-iii) Charge density plots of plasmon-like states of the $Au_{100}$ nanorod corresponding to two close absorption peaks around 2 eV, which upon excitation quickly (< 30 fs) lead to the population of Au bulk-states (iv) from which electron transfer to the acceptor state of $MoS_2$ (v) can then take place.[209] Reprinted with permission from Elsevier.

The faster decay component was assigned to the electron-phonon coupling, which varied from a few hundred fs to ps, depending on the adsorbed energy density by the NPs, and a slower decay component, assigned to ph-ph coupling.[8,232] It was shown that the more power is adsorbed by the NPs, the more energetic are the plasmon generated hot carriers, and consequently the longer it takes for them to lose their energy by e-ph coupling. By extrapolating the e-ph coupling time to zero adsorbed power, the characteristic e-ph coupling times were extracted, independent of the adsorbed power. This quantity depends only on the material properties (electron-phonon coupling constant and the electronic heat capacity) thus it can be used to determine the influence of the composition material on the ultrafast dynamics of hot carriers. For AuNPs, an e-ph coupling time of 0.9 ps was found, whereas for both Au/Pt and Au/Ni NPs, a time of 0.4 ps was determined. The slower, ph-ph coupling time was found to be 629 ps for AuNPs, 782 and 349 ps for Au/Pt and Au/Ni nanoparticles, respectively. This longer relaxation time reflects the cooling of nanoparticles through heat diffusion at the surface of the NPs and to the substrate, thus it is intimately linked to the interfacial thermal resistance of each NP.

An important advantage of plasmonic photocatalysts is the dynamical tunability of their optical response not only based on the shape of individual NPs, but also in 2D or 3D collections of plasmonic NPs (i.e., metamaterials).[233] In a remarkable study, Harutyunyan *et. al*. showed the influence of hot spots (tight spaces in between adjacent nanostructures with enhanced EM fields) in the spectral and time domain response of plasmon resonance and hot carriers dynamics.[234] As opposed to modifying the plasmonic photocatalyst and form hybrid structures for example, the authors changed the geometry of the plasmonic photocatalysts. They used Au disks deposited on an Au surface, with alumina or $TiO_2$ spacers of varying thickness in between the two metal surfaces. Based on the separation between the gold surfaces (i.e., the gap of the hot-spot), and on the spacer composition, the authors observed anomalously strong changes to the ultrafast temporal and spectral responses of the plasmon generated hot carriers (see **Fig. 5c**). The authors showed a large ultrafast (pulse-width-limited) contribution to the hot carriers' decay signal in nanostructures containing hot spots, which correlates with the efficiency of the generation of highly excited surface electrons. This effect was attributed to the generation of hot carriers from hot spots, with a much higher generation rate than outside those hot points, thus suggesting that plasmonic nanostructures featuring hot spots that are accessible to molecules, should be cleverly designed for an efficient photocatalytic activity.

Alongside the observed evidence through experimental studies, the clarification of the mechanisms of reactions catalyzed by hot carriers is a rather complex task. It has been shown that local thermal effects, hot carriers injection and electromagnetic coupling in confined space can all actively contribute to the modification of reaction yields and rates on the different timescales considered here.[83,84,235,236] Theoretical modelling has the potential to provide original insights. To give some examples, by employing coupled Boltzmann-heat equations and Arrhenius law-based models it was recently suggested that plasmon-induced local heating of the environment may significantly affect chemical reaction rates.[237–240] As remarkable example, Huang *et al*. and Yan *et al*. investigated the water splitting reaction close to Au plasmonic nanoclusters by employing a nuclear-electron quantum dynamics approach based on rt-TDDFT coupled to Ehrenfest dynamics, showing that both the plasmon-induced local field enhancement and a direct electron transfer to an antibonding orbital of a water molecule, due to de-localized excitations, significantly affect the photoreaction dynamics (see **Fig. 6**).[241,242] Noteworthy, modelling has also highlighted situations where hot carriers are not directly involved in plasmon-enhanced reaction. Spata and Carter, by means of high-level quantum mechanical calculations using the ECW embedding scheme, showed that the H$_2$ desorption from a Pd catalytic reactive centre can be favoured by coupling the nano system with a plasmonic Al NP thanks to a purely electromagnetic enhancement effect.[243]

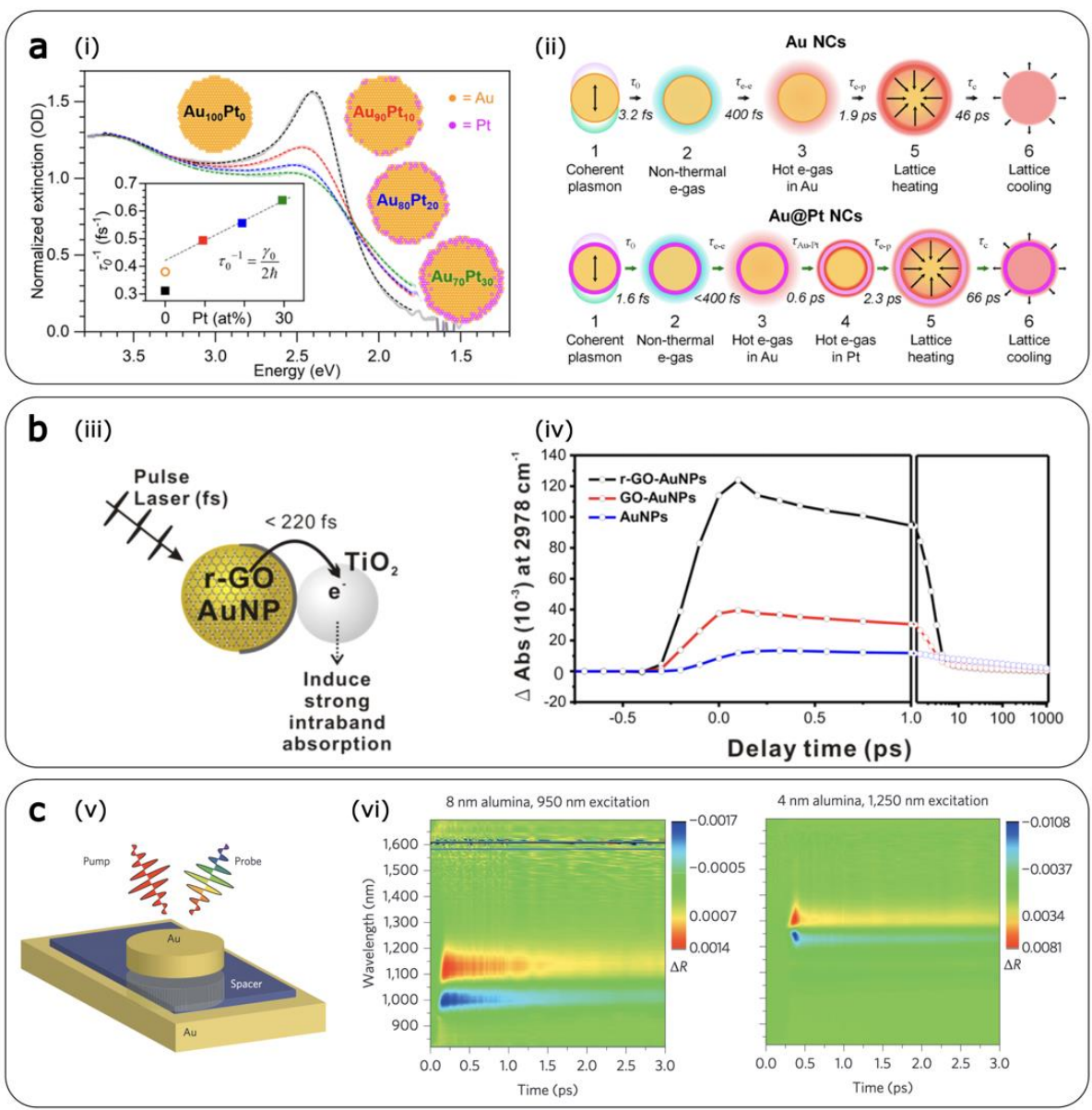

**Figure 5. Ultrafast plasmon generated hot carriers' dynamics and their interaction with adsorbed materials.** (a) (i) The broadening of the LSPR of gold nanoparticles (AuNPs) following the deposition of Pt layers. The inset shows the plasmon dephasing rate, extracted from the spectral domain optical response, in AuNPs (orange circle) and of Au@Pt core-shell NPs. (ii) Time dynamics of surface plasmon resonance excitation and dephasing, in Au NPs and Au@Pt core-shell NPs, respectively.[226] Reprinted with permission from ACS. b) (iii) The ultrafast charge transfer from reduced graphene oxide AuNP (r-GO-AuNP) to adsorbed TiO$_2$. (iv) The change of the charge transfer efficiency from Au NPs to TiO$_2$ due to the deposition of GO and r-GO respectively on AuNPs. The GO monolayer improves the mobility of hot carriers.[227] Reprinted with permission from RSC. (c) (v) Experimental setup for monitoring the influence of hot spots on the ultrafast response of plasmon generated hot carriers. (vi) By decreasing the alumina spacer thickness from 8 to 4 nm, an anomalous ultrafast response of the plasmon generated hot carriers appears due to the hot spot.[234] Reprinted with permission from ACS.

On similar grounds, ammonia $NH_3$ decomposition on Ruthenium-doped Copper NPs[244] and carbon-fluorine bond activation of $CH_3F$ in presence of Al-Pd plasmonic nanostructures[245] were both shown to be enhanced under illumination. In both cases, the analysis of the excited state minimum energy path (MEP) computed through ECW revealed lower activation barriers compared to the ground state pathways, thus justifying the speed-up of the reaction rates under light irradiation. In those cases, the plasmon can enhance the population of excited state species[243,244], possibly induce a decrease of the energy barriers because of hot carriers related effects[245], or even a simultaneous combination of both.

Concluding this Section, combining experimental and theoretical approaches to study ultrafast phenomena in (hybrid) plasmonic photocatalytic materials is an invaluable and desirable tool to catch a glimpse of the complex fundamental processes involved. As a last example, ultrafast pump-probe optical measurements have unraveled the energy flow in hybrid plasmon photocatalysts. Although it was traditionally assumed that plasmon-generated hot carriers are formed homogeneously throughout the volume of the plasmonic material, recent ultrafast studies have shown that in hybrid plasmonic materials (such as $Ag/TiO_2$, Au/Pt) the energetic charge carriers are formed predominantly at the interface of the two materials, due to the interfacial energy states.[246–248] This allows efficient charge transfer to adsorbed molecules, driving chemical reactions. This paradigm-shift in plasmonic photocatalysts, enabled partly by ultrafast optics, opens the road to limit increasingly more the losses associated with plasmonic chemistry, and the design of novel plasmon-driven photocatalysts.[249–251]

## 4. Ultrafast multi-functional plasmonics

During the last decade, a new subfield of plasmonics, the so-called *active* plasmonics,[252–254] has emerged, with the goal of tuning plasmons using an external stimulus such as an electric or magnetic field, or even a mechanical stress/acoustic excitation. As noticeable example, instead of the conventional noble metals, the use of magnetic or strongly correlated materials allows for an additional degree of freedom in controlling EM field properties.[255,256] In fact, these materials enable light to interact with the spin of electrons. Thus, light can be used to actively manipulate the magnetic properties of such materials.[257,258] Plasmonic and magnetic properties can be combined in these material at the nanoscale through various magneto-optical effects.[66,255,256] Furthemore, plasmonic modulation via lattice vibrations is an additional method for active control of SPs.[259] As a result, multifunctional plasmonic materials that allow for dynamic manipulation of light properties (e.g., amplitude and polarization) at the nanoscale are critical features for future all-optical nanotechnologies based on active photonics. This section provides an overview of recent studies of ultrafast dynamics in plasmonic nanostructures with magnetic, acoustic, and plasmonic functionalities, as well as how these functions can be combined to obtain novel ultrafast phenomena at the nanoscale.

Strategies for ultrafast all-optical control of magnetism have been the subject of intensive research for several decades due to the potential impact on technologies such as magnetic memories[260] and spintronics,[261] as well as the opportunities for nonlinear optical control and modulation[262] in applications such as optical isolation and non-reciprocity.[263]

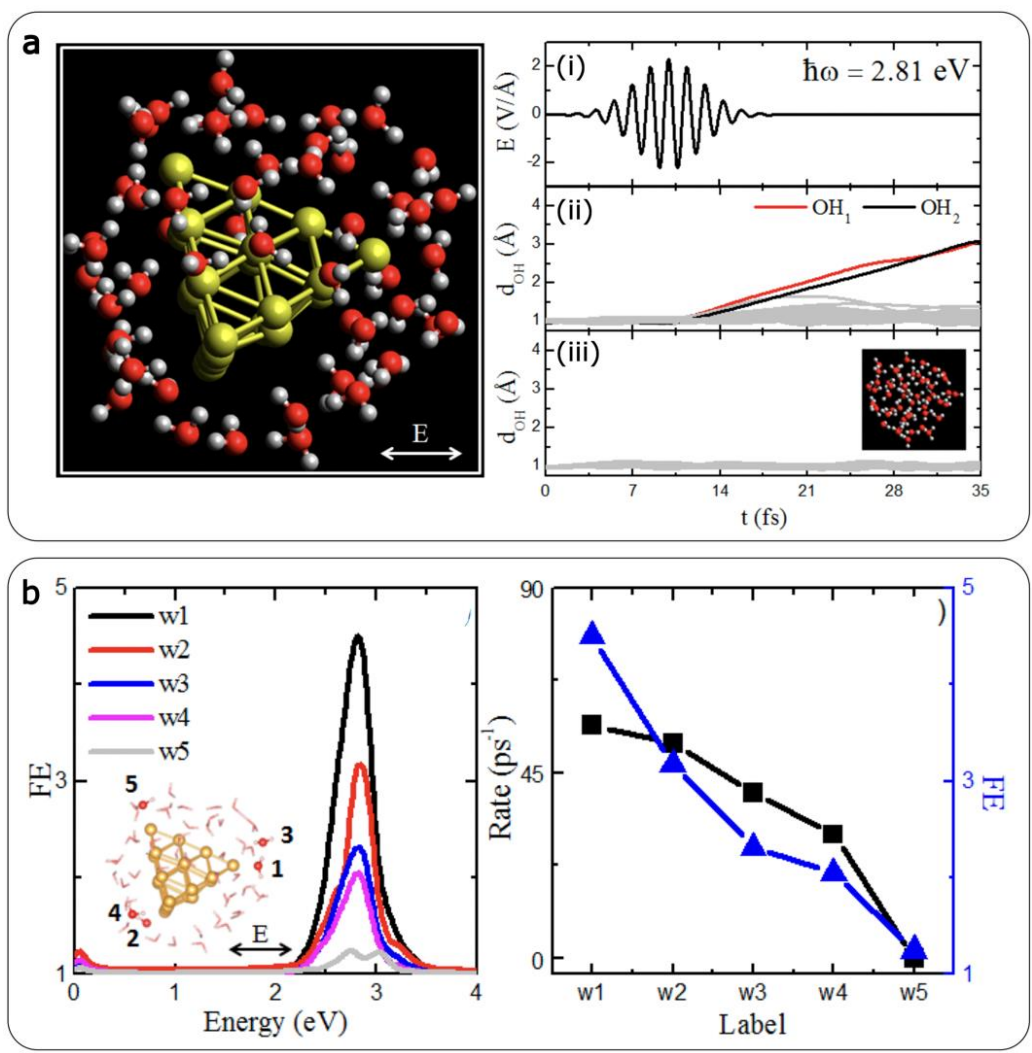

**Figure 6. Theoretical simulations of water splitting reaction close to a plasmonic Au$_{20}$ cluster employing rt-TDDFT coupled to Eherenfest dynamics**. (a) Snapshot of the Au$_{20}$ cluster surrounded by water molecules (left panel). The white arrow displays the polarization of the incoming electric field whose frequency and time evolution are displayed on the right (i). Interestingly, upon excitation and in presence of the Au cluster, some OH bonds break down, as shown by the time evolution of the OH bond lengths of all the water molecules surrounding the cluster (ii), thus triggering the water splitting reaction. Conversely, without the Au cluster (iii), all the OH bonds oscillate over time, remaining intact. (b) Local field enhancement (FE) as a function of excitation energy in different spatial positions (corresponding to different water molecules) surrounding the Au cluster (left panel). The corresponding rate of water splitting evaluated at the same spatial points (w1-5) is shown on the right. Notably, the rate of water splitting does not follow exactly the field enhancement trend, indeed even direct charge transfer excitations turned out to affect the reaction rate for some water molecules depending on their distance to the Au surface.[242] Reprinted with permission from ACS.

One of the most intriguing challenges in this field concerns the maximum modulation speed of nanophotonic devices and the time limit for magnetization reversal (switching) in magnetoplasmonic and, more in general, in magneto-optical devices. In the case of plasmonic structures, this modulation speed depends on the e-ph relaxation time, which in metals is a few ps, resulting in GHz range bandwidth. Several strategies have been proposed to overcome this limitation, for instance by exploiting nonlinear optical effects. In such processes, no direct transfer of carriers into excited states occurs, that is there is no time limitation for the excitation or relaxation processes, and therefore, the phenomenon can take place in a virtually instantaneous manner and it depends only on the duration of the light pulse exciting the structure.[264] This possibility will be discussed in more detail in **Section 5**. Here, we focus on discussing the opportunity to achieve ultrafast switching applications in the framework of magnetic materials, such as Ni, which can be combined with plasmonic structures[265–267] or be used directly as plasmonic building blocks.[268–272]

As highlighted many times in the previous sections, plasmonics enables confinement and enhancement of EM radiation well below the diffraction limit. This is a crucial feature for opto-magnetic applications in spintronics, where deterministic control of nanometer-sized magnetic bits is a major objective. Incorporating plasmonic nanostructures into magneto-optical active materials dramatically improves pump-induced control of the magnetization state at the ultrafast timescales. The fastest and least dissipative way of switching the spin of electrons is to trigger an all-coherent precession. Schlauderer *et al.* showed that THz EM pulses allow coherent steering of spins by coupling the antiferromagnetic material $TmFeO_3$ (thulium orthoferrite) with the locally enhanced electric field of custom-tailored antennas (see **Figure 7a**).[273] Similarly, but using visible (PHz) radiation, Mishra *et al.* employed resonant EM energy funneling through plasmon nanoantennas to influence the demagnetization dynamics of a ferrimagnetic TbCo thin films (see **Fig. 7b**). They demonstrated how Ag nanoring-shaped antennas under resonant optical fs pumping reduce the overall demagnetization in the underlying films up to three times compared to non-resonant illumination, and attributed such a substantial reduction to the nanoscale confinement of the demagnetization process.[65] Very recently, they have also devised a new architecture simultaneously enabling light-driven bit downscaling, reduction of the required energy for magnetic memory writing, and a subtle control over the degree of demagnetization. To achieve these features, they employed in a magnetophotonic surface crystal featuring a regular array of truncated-nanocone-shaped Au-TbCo antennas showing both localized plasmon and surface lattice resonance modes.[274]

Another interesting approach to achieve ultrafast control of magnetism at the nanoscale is represented by the all-optical switching (AOS) mechanism, where the optical laser pulses are used to deterministically reverse the spin moment.[275–277] This discovery could have important technological implications, since, for example, switching the spin magnetization by an ultrashort pulse could lead to much faster writing of magnetic bits in magnetic recording media. The origin of the AOS was initially thought to be linked to the helicity (right or left circular polarization state of the light pulse), i.e., the injected "photonic spin moment" (the so-called spin angular momentum, SAM, of light), but subsequent investigations showed that solely fast laser heating was sufficient to trigger the magnetization reversal.[278] By using a microscope objective[279] or nanopatterning of magnetic materials[280–282], bit size can be reduced to a few hundred nanometers.

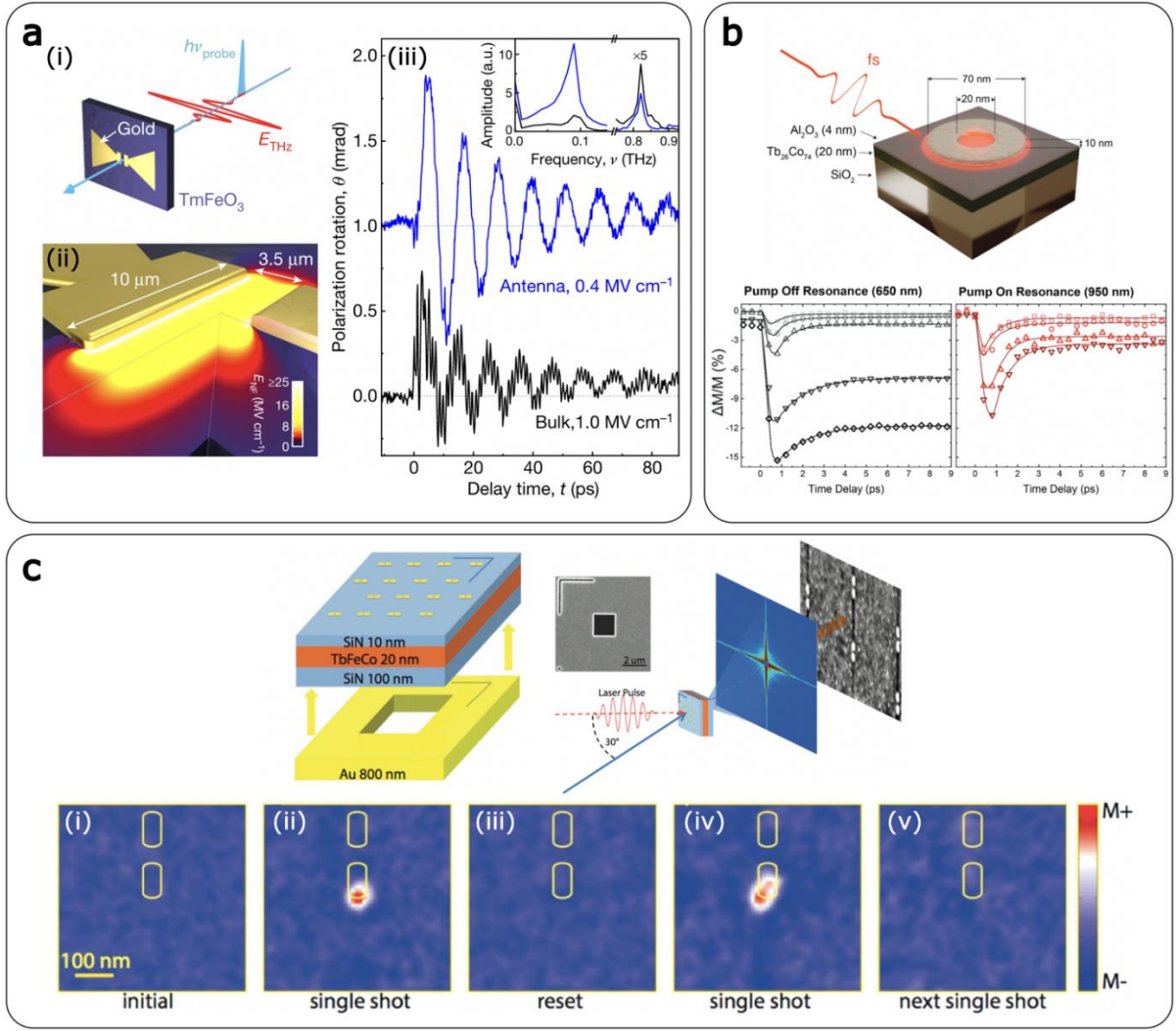

**Figure 7. All-optical plasmon-driven magnetization dynamics.** (a) (i) Schematic of the gold bowtie antenna on TmFeO$_3$. The structure is excited from the back side by an intense THz electric field E$_{THz}$ (red waveform) while a co-propagating near-infrared pulse (light blue) probes the induced magnetization dynamics in the feed gap between the two lobes of the antenna. (ii) Peak near-field amplitude in the antenna feed gap calculated by FDTD simulations for an incident THz waveform with a peak field amplitude of 1 MV cm$^{-1}$. (iii) Experimentally detected polarization rotation signal obtained for a peak electric THz field of 1 MV cm$^{-1}$ on the unstructured substrate (black curve), and when probing the feed gap of the bowtie antenna structure resonantly using a THz waveform with a peak electric field amplitude of 0.4 MV cm$^{-1}$ (blue curve). Inset: corresponding amplitude spectra featuring two modes (at 0.09 THz and 0.82 THz).[273] Reprinted with permission from Springer-Nature. (b) Schematic of Ag nanoring on Tb$_{26}$Co$_{74}$. Pump-probe measurements was performed to determine the magnetization dynamics. The differential magnetization using an off-resonance (on-resonance) pump is displayed at the bottom left (right), showing a great enhancement of the demagnetization by the on-resonance pump.[65] Reprinted with permission from RSC. (c) The schematic show TbFeCo film with gold two-wire antennas exited by a laser pulse. AOS was achieved with a size of 53 nm using the nanoantennae to confine fields spatially. X-ray holography is used to image the magnetic switching. The switching is shown to be both reproducible (i-iv) and reversible (v).[283] Reprinted with permission from ACS.

In a seminal work, Liu *et al.* exploited LSPs in gold dimer-nanobar nanoantennas on a TbFeCo alloy thin film to achieve a 50 nm AOS-switched spot size[283] (see **Fig. 7c**). Thus, plasmonics can enable AOS to become technologically relevant, since plasmons can allow to reach bit sizes which can be used in heat-assisted magnetic recording[284] and allow deterministic ultrafast all-optical magnetization switching at the nanoscale. Moreover, strong localization of the pump pulse by SP excitation reduces the laser fluence required for the ultrafast demagnetization.[285,286]

One of the possible ways for the photon to act on the spin moment could be through an opto-magnetic effect, the inverse Faraday effect (IFE). This non-linear opto-magnetic effect, discovered in the 1960s, describes the generation of an induced magnetic moment by a circularly polarized EM wave.[287] To explain the IFE in bulk materials, several models have been proposed recently. The deeper understanding of the origin of the IFE is still the subject of ongoing investigations, and here we report the most relevant works in relation to the topic of this review. Hertel developed a plasma model[288], which was recently used by Nadarajah and Sheldon to estimate the magnetic moment that could be induced in a plasmonic Au nanoparticle.[289] Popova *et al.* employed a four-level hydrogen model with impulsive Raman scattering to show that such a process could lead to a net induced magnetization.[290] Using relativistic electrodynamics, Mondal *et al.* showed that there exists an EM wave–electron spin interaction that provides a linear coupling of the photon SAM to the electron spin, which then acts as an optomagnetic field that generates the induced magnetization.[291] By using a second-order density matrix perturbation theory a quantum theory for the IFE was derived, by taking into account also the optical absorption.[292,293] Carrying out *ab-initio* calculations for bulk Au, Berritta *et al.* computed a moment of $7.5 \times 10^{-3}$ $\mu_B$, where $\mu_B$ is the Bohr magneton, per Au atom by pumping with circular EM radiation with a 10 GW/cm$^2$ intensity at 800 nm.[292] Detailed measurements of the light-induced magnetization in Au, however, are still very few. In an early pioneering investigation, Zheludev et al. could measure an induced polarization rotation of $\sim 7 \times 10^{-4}$° in an Au film upon pumping with a laser intensity of 1 GW/cm$^2$ and 1260 nm wavelength, a value that is within an order of magnitude consistent with the *ab-initio* calculated induced moment.[294] Recently, Cheng *et al.* reported the experimental optically induced magnetization in plasmonic gold nanoparticles, with magnetization and demagnetization kinetics that are instantaneous within the sub-ps time resolution (see **Fig. 8a**).[295] Under typical ultrafast pulse excitation (>10 GW/cm$^2$ peak intensity), the induced magnetic moment $2.9 \times 10^7$ $\mu_B$ or 0.95 $\mu_B$ per Au atom, i.e., two orders of magnitude higher than that predicted by theory for bulk Au.[292]

All the above considerations were built upon exploiting the SAM of the EM field. An optical beam can also carry a well-defined orbital angular momentum (OAM).[296,297] Combining OAM with plasmonics, it has been demonstrated that sub-fs dynamics of OAM can be realized in nanoplasmonic vortices.[88,298,299] Hence, plasmonic vortices carrying OAM can be confined to deep subwavelength spatial dimension and could offer an excellent time resolution. The OAM can, therefore, be expected to enter soon the developing area of magnetophotonics, where the OAM could offer a new functionality to control the nanoscale magnetism.[299] Although there is an emerging understanding of the IFE coupling of the SAM of a beam to the electron spin, a similar understanding of the interaction of OAM with spin or orbital magnetism has still to be established. Recently, observations of interaction of magnetism and an OAM vortex beams were reported.[300,301] Karakhanyan *et. al* quantified numerically the relative contributions of SAM and OAM of light to the IFE in a thin gold film illuminated by different focused beams carrying SAM and/or OAM, by using a hydrodynamic model of the conduction electron gas.[302] The OAM of light provides a new degree of freedom in the control of the IFE and the resulting opto-magnetic field, which has the potential to influence numerous research fields, such as all-optical magnetization switching and spin-wave excitation. These findings support a mechanism of coherent transfer of angular momentum from the optical field to the electron gas, and they pave the way for optical subwavelength strategies for optical isolation that do not require externally applied magnetic fields. It can already be perceived that taking both spin and orbital degrees of freedom of photonic beams into account will become paramount for the future development of ultrafast magnetophotonics.

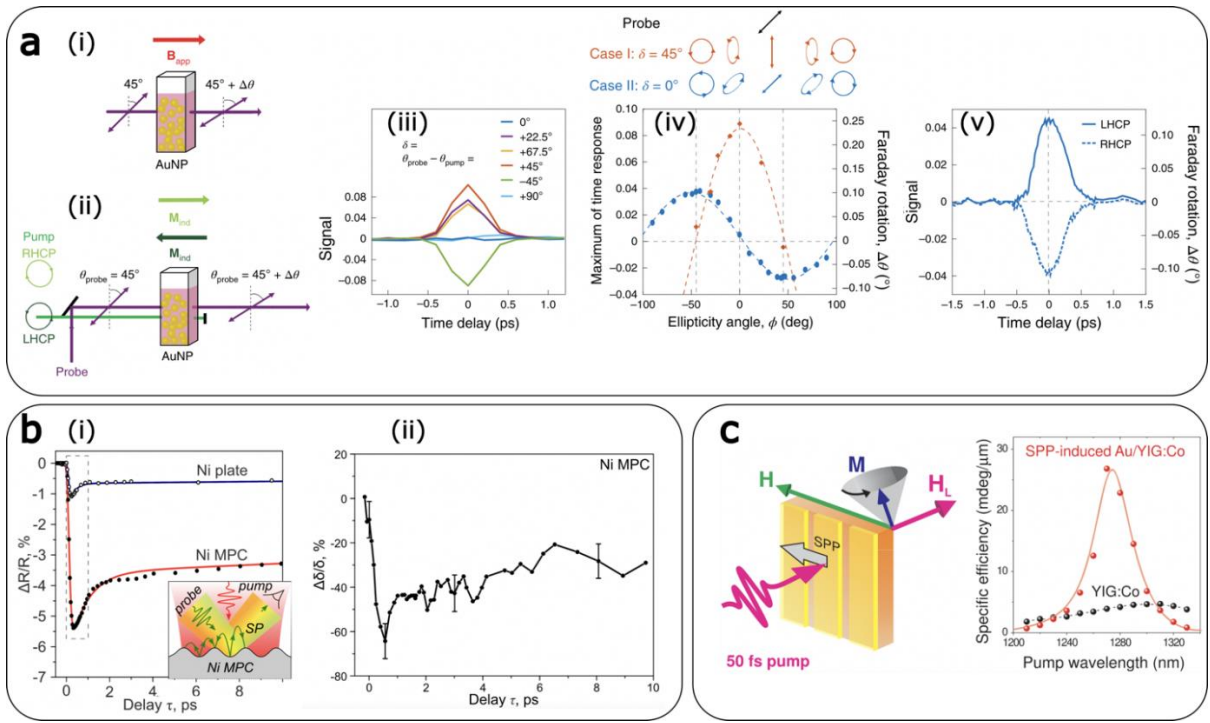

**Figure 8. Ultrafast magneto-plasmonics.** (a) (i) The Faraday effect is the rotation of the polarization plane of light transmitted through a magnetized medium. (ii) The IFE is the induced magnetization of a medium, $M_{ind}$, during circularly polarized excitation (green line). The direction of induced magnetization depends on the helicity of the light. The optical rotation of a probe beam (purple line) indicates $M_{ind}$, and $B_{app}$ is the external applied magnetic field; $\Delta\theta$, Faraday rotation angle; R(L)HCP, right-(left-)handed circularly polarized light. (iii) Light-induced rotation due to only the optical Kerr effect ($\phi = 0°$), as a function of $\delta$. (iv) Light-induced rotation as a function of $\phi$. Red dots represent data from experiments with the geometry of case I, where $\delta = 45°$. Blue dots represent data for the case II geometry, where $\delta = 0°$. Dashed lines are fits to the data. (v) Light-induced rotation due to pure IFE when the pump is circularly polarized with left-handed helicity ($\phi = -45°$, solid line) or right-handed helicity ($\phi = +45°$, dashed line),

corresponding to the positive and negative maxima for case II in (iv).[295] Reprinted with permission from Springer-Nature. (b) (i) Inset: Model visualization of ultrafast all-optical modulation of SP-assisted TMOKE in a one-dimensional (1D) nickel magnetoplasmonic crystal (MPC). The differential reflectance is displayed for plain nickel and for the nickel MPC. (ii) Laser induced modulation of the SP-enhanced TMOKE in 1D nickel MPC as a function of the pump−probe delay.[63] Reprinted with permission from ACS. (c) SP induced amplification of photomagnetic spin precession. A YIG:Co film with Au gratings, placed in an external in-plane magnetic field, that was excited by a 50 fs pump (left) displayed an increase in its specific efficiency of the spin precession (right), as opposed to such a film without the Au. Dots represent experimental data and the full line is results from a numerical simulation.[303] Reprinted with permission from ACS.

Plasmon-enhanced magneto-optical effects can allow to improve by orders of magnitude the detection sensitivity of ultrafast magnetic processes. Novikov *et.al.*[63] investigated experimentally the ultrafast modulation of the SP-resonant transverse magneto-optical Kerr effect (TMOKE) induced by a non-resonant 50 fs pump laser pulse in the one-dimensional magnetoplasmonic crystal (see **Fig. 8b**). They observed the suppression of TMOKE in the SP-resonant probe from 1.15% to 0.4% by the demagnetization of the nickel MPC under a nonresonant pump. The absolute TMOKE difference was around 20-fold enhanced in the MPC compared to the plane nickel film, thus proving the higher sensitivity of plasmonic systems to magnetization dynamics. Furthermore, they demonstrated that the carrier dynamics induced by the ultrashort light pulse affected the SP excitation process itself: the laser heating induces the modification of the SP wavevector through the modulation of the metal optical constants. As a result, a differential reflectivity value as high as 5.5% is achieved in MPC in comparison with 1.1% value in the plane nickel. Finally, it was shown that electron thermalization and relaxation dynamics are slower in MPC relative to the plane nickel, since e-ph relaxation times extracted from the differential reflectance were 800 and 260 fs for the MPC and plane nickel film, respectively. Similarly, Kazlou *et al.* recently demonstrated SP-assisted control of spin precession phase in hybrid noble metal–dielectric magneto-plasmonic crystals (see **Fig. 8c**).[303,304] These results can be applied to the development of new magnetoplasmonic devices, providing more sensitivity to the magnetic order on the sub-ps timescale, where non-thermal effects occur. In this framework, particularly interesting would be the study of plasmon dephasing (happening on the time scale of few tens of fs) on the spin dynamics in opto-magnetic-active nanomaterials.[66] By exploring non-thermal pathways, where the intrinsic losses due to heating can be overcome by exploiting the dynamics of the electrons on sub-100 fs timescales, can open excellent perspectives in both the fundamental and applied aspect of ultrafast magnetoplasmonics.

As mentioned at the beginning of this Section, plasmonic modulation by means of ultrafast lattice vibrations represents an interesting approach for the active control of SPs, as lattice contraction can red-shift the SP resonance frequency.[305] In 1988 van Exter and Lagendijk showed that SP excitation in silver in the Kretschmann geometry can significantly amplify pump-probe signals compared to standard reflectivity measurements.[306] An acoustic variation of the order of 4% in the lattice density altered the wave vector of the time-delayed SP probe pulses, causing a significant shift in probe reflectivity in the Kretschmann configuration. They concluded that the intrinsic lifespan of longitudinal phonons in silver at 42 GHz surpassed 100 ps based on the decay duration of these thermo-acoustic oscillations, which was mostly governed by the finite (52%) acoustic reflection at the silver-glass interface. Measurements of the mean free path (or lifetime) of high-frequency phonons in solids, caused by the interaction between ballistic and diffusive heat transport, are crucial for understanding the nanoscale thermal characteristics of materials.[307]

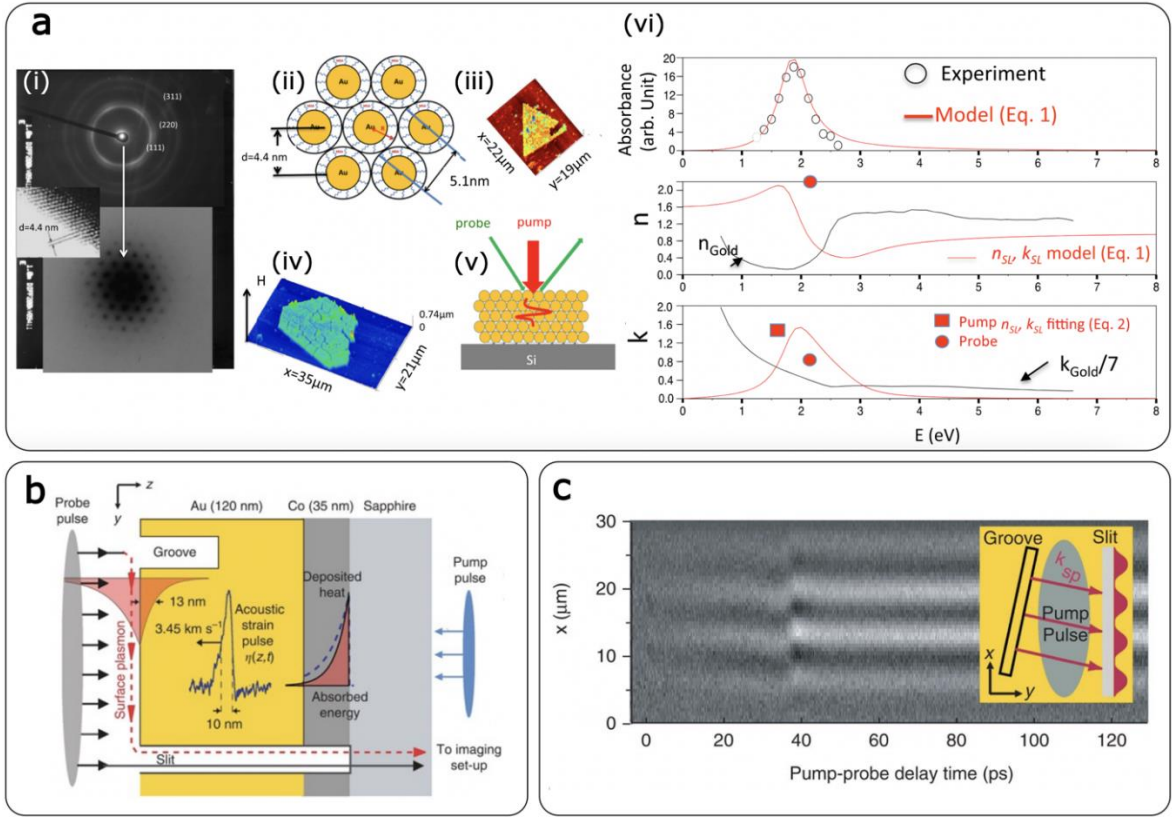

**Figure 9. Ultrafast acousto-plasmonics.** (a) (i) Transmission electron diffraction (TED) pattern showing the Debye Scherrer rings (top figure) coming from the crystallized cubic gold nanoparticles. In the middle figure, the transmission electron microscopy image (TEM) shows the mesoscopic hexagonal arrangement (hcp) with the two-neighbor plane distance of 4.4 nm. The hcp arrangement is also well evidenced with small angle electron diffraction revealing the 6-fold axis (bottom figure). (ii) Sketch of the gold nanoparticles superlattices connected via mercaptosuccinic acid molecules. (iii) Optical microscopy view of a superlattice deposited onto a silicon substrate. (iv) AFM profile of the hexagonal shape superlattice (height H=180 nm). (v) Sketch of the pump and probe experiments performed on these nanoparticles superlattice where coherent acoustic phonons are generated and detected in the front-front configuration. (vi) Top panel: plasmon resonance of the gold superlattice (experimental data, dots, and calculated, red line). Middle and bottom panels: refractive index (real n and imaginary parts k) of the gold superlattice compared to those of bulk gold. The red symbols (square and circles) are the refractive index values estimated with the photoacoustic response. [308] Reprinted with permission from APS. (b) Schematic drawing of the acousto-plasmonic pump-probe experiment: surface plasmons propagating at the gold–air interface probe the reflection of acoustic pulses generated in the laser-heated cobalt transducer. (c) Measured acousto-plasmonic pump-probe interferogram showing a pronounced shift of the interference fringes upon reflection of an ultrashort acoustic pulse. The inset illustrates the geometry of the plasmonic slit-groove interferometer[309]. Reprinted with permission from Optica Publishing Group.

Similar time-resolved experiments in thin gold and copper films[310,311] have revealed the contribution of short-lived nonequilibrium electrons to the buildup of mechanical stress driving the initial stage of thermo-acoustic expansion, which is consistent with similar measurements in plasmonic nanoparticles[312]. For example, Deacon *et al.* demonstrated that a highly localized acoustic resonance exists in the nanoparticle-on-mirror structure, near the nanoparticle's narrow base.[313] This bouncing mode has an extremely variable period pertaining to the nanoscale morphology attainable through matched plasmonic coupling by means of a coupled mode whose optical field is likewise tightly contained within the nanoscale void beneath the facet. Due to this restriction on both, the plasmonic and acoustic modes are coupled to the gap by the significantly improved acousto-optic coupling as is observed in comparison to conventional acousto-optic crystals, impacting the SP resonance. Ruello *et al.* measured coherent GHz acoustic phonons in plasmonic gold nanoparticle superlattices (NPSs), with a typical in-depth spatial extension of approximately 45 nm, which is approximately four times the optical skin depth in gold (see **Fig. 9a**).[308] The modeling of transient optical reflectivity suggested that phonons are generated by ultrafast heating of the NPs aided by light activation of the volume plasmon polariton. Based on these findings, it has been shown that it is possible to map the photon-electron-phonon interaction in subwavelength nanostructures,[314–316] which provides interesting insights into the fundamental features of these architectures. Finally, it is worth mentioning that an interesting approach to further boost research in ultrafast active plasmonics is represented by the combination of acoustic, magnetic and plasmonic properties. In this context hybrid gold-cobalt bilayer structures[252] can be suitable candidates for opto-magneto-acoustic switching experiments (see **Fig. 9b,c**).[309,317]

## 5. All-optical switching applications and new materials

One of the main challenges in all-optical information processing is the creation of an efficient and compact high-speed all-optical switch, the equivalent of the transistor in conventional electronics. For practical applications in integrated photonics, all-optical modulators should ideally feature sub-ps switching times at fJ control light energies, as well as switching contrasts greater than 10 dB and characteristic sizes below 100 nm. Up to date, no engineered photonic structure that can operate in the THz bandwidth range fulfills all these conditions. In this context, plasmonics emerges as a suitable approach for all-optical processes as plasmons can rapidly alter the complex refractive index of a medium with an enhanced response[318–325]. Upon illumination of a metal by a light pulse (pump), non-thermalized electrons are excited, which subsequently decay through several processes, spanning different timescales, from the fs up to ns range, leading to dynamic changes of the material permittivity. In the context of all-optical switching, a control light pulse is used to vary transmission or reflection between high and low states, to modulate the intensity of a signal pulse which carries the information. Given that plasmonic resonators can spectrally tailor light dispersion, wavelength regions showing negligible levels of "slow" temporal components connected to e-ph and ph-ph scattering processes can be attained, where only sub-ps responses dominate. By engineering nanostructured gold, this approach has demonstrated differential transmission modulations up to > 20% with temporal widths < 300 fs at ~1 mJ/cm$^2$ pump fluence, as shown in **Fig. 10a**.[325] Both positive and negative transmission changes can be observed, depending on the probe wavelength, as enabled by the pump-induced shift of the resonance spectrum. Considering the high electron densities present in noble metals, their electron distributions cannot be substantially modified optically, preventing larger permittivity modulations[262]. In this scenario, indium tin oxide (ITO) arises as a promising material that exhibits significantly smaller - but sufficiently high – excited electron concentrations in the near infrared.[326,327] Moreover, ITO also possesses an epsilon-near-zero (ENZ) region, characterized by a permittivity spectrum with a zero crossing of the real component, where optically induced variations of the refractive index produce the strongest effects on resonant features.[262,328–330] Coupling a thin ITO layer at its ENZ region to an array of gold nanoantennas (see **Fig. 10b**) to modulate the plasmonic resonance has revealed absolute transmission changes around 20% at a pump fluence of only < 1 mJ/cm$^2$ (equivalent to ~0.3 pJ per nanoantenna's geometric cross-sectional area).[329] The fastest reported plasmonic switch has been achieved by exploiting gold's high third order nonlinearity through two-photon absorption (TPA).[331] TPA influences the imaginary part of the refractive index of a medium in a nearly instantaneous manner. In a pump-probe configuration, the transmission of the probe beam changes abruptly whenever a pair of pump and probe photons are absorbed simultaneously, recovering its nominal value as soon as the pulses no longer temporally overlap. The fundamental time limit for this process is practically either the pulse duration or the resonance bandwidth. The differential transmission temporal trace of a Fano-resonant gold metasurface shown in **Fig. 10c**, displays a temporal width of only 40 fs, with a modulation depth of ~10% at 0.07 mJ/cm$^2$ pump fluence (~0.1 pJ per unit cell), which reaches 40% at 0.3 mJ/cm$^2$ (~0.5 pJ per unit cell)[331]. Similar performances has also been measured in nonlinear dielectric nanostructures at Mie resonances (see **Fig. 10d**).[264,332,333] Outstanding differential (absolute) transmission modulations greater than 100% (20%) at ~4 mJ/cm$^2$ pump fluence, were also reported in ITO nanowires (see **Fig. 10e**).[327,334]

Another avenue to control the strength of light–matter coupling to exceed losses could combine optical/plasmonic field properties of a resonator with the electronic excitations of an active material.[335,336] Layered materials with a 2D character such as graphene and transition metal dichalcogenides (TMDs) have also been considered for all-optical processing devices,[337–339] as they support strong light-matter interactions and display optical nonlinearities enabling both atomic-scale and a sub-fs modulation of light properties.[340–342] In the past years, research efforts have been spent to study ultrafast nonlinear optical processes in layered materials, in particular TMDs.[104,343–350] The most efficient nanoscale all-optical switch has been obtained by using graphene as the nonlinear element.[337] Graphene is a nonlinear saturable absorber, meaning that its transparency increases under intense light excitation due to photogenerated carriers causing Pauli-blocking, with relaxation times in the sub-ps range. To further enhance light-matter interactions, M. Ono et al.[337] placed a graphene bilayer on top of a plasmonic waveguide with a 30 nm slot width (see **Fig. 10f**), significantly reducing the required power for saturable absorption. They demonstrated a differential transmission increase of >100% at just 35 fJ switching energy, with a 260-fs temporal response. To reach sub-fJ switching energies, Nozaki et al. used a combination of a photonic-crystal nanocavity and strong carrier-induced nonlinearity in InGaAsP to demonstrate low-energy switching within a few tens of ps. Switching energies with a contrast of 3 and 10 dB of 0.42 and 0.66 fJ, respectively, have been obtained.[351]

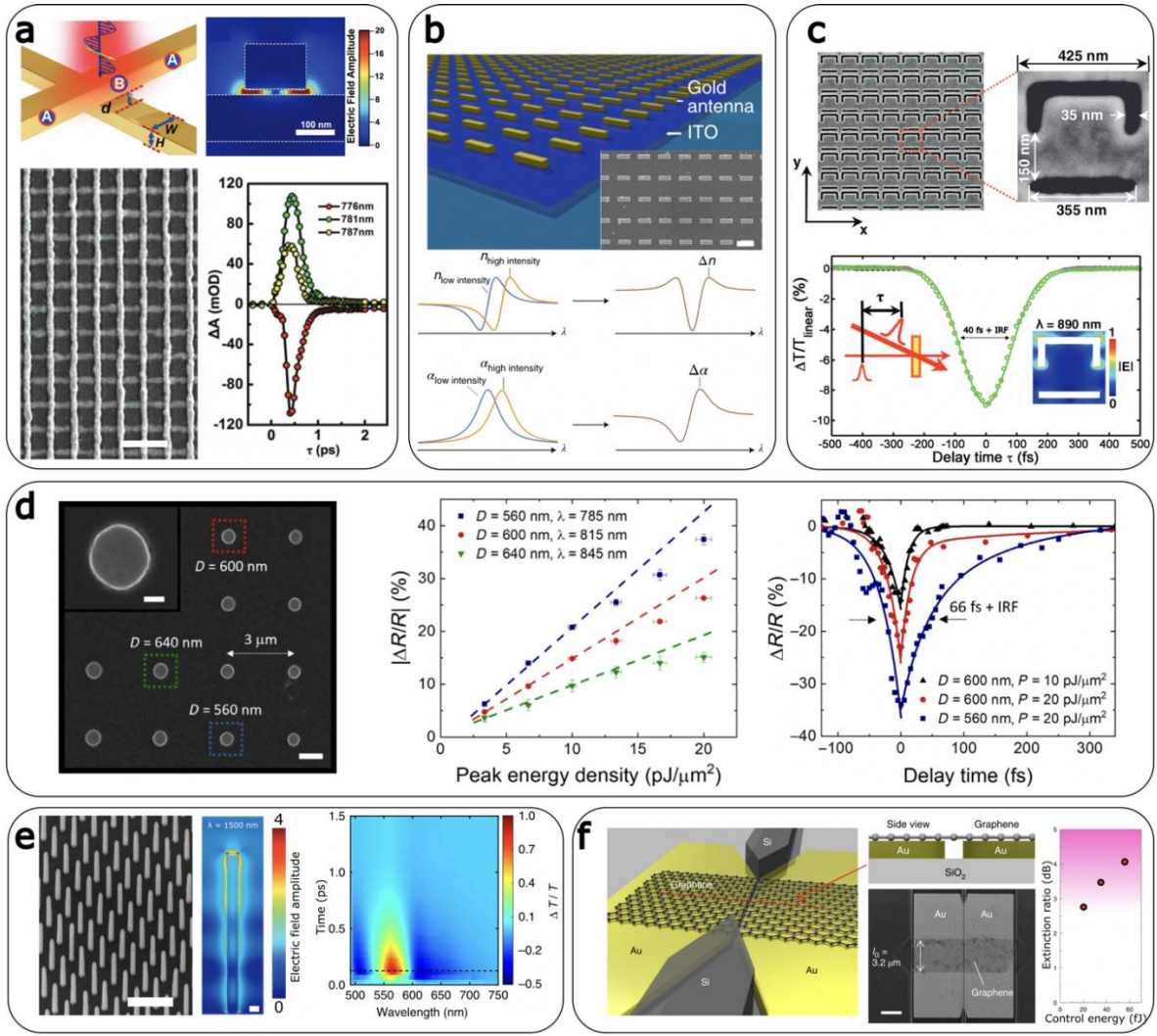

**Figure 10. Ultrafast plasmonic all-optical switching.** (a) Cross-stacked gold nanowire network supporting a Fano resonance. Design, SEM image (scale bar, 1 μm), electric field enhancement distribution and ultrafast performance as measured through pump-probe spectroscopy (±110 mOD corresponds to ±22% transmission modulation).[325] Reprinted with permission from RSC. (b) Array of gold nanoantennas coupled to an ITO layer (SEM image in the inset; scale bar, 500 nm). A small variation in ITO's refractive index causes a significant spectral shift of the resonance wavelength of the nanoantennas, causing a dramatic change in the effective refractive index ($\Delta n$) and absorption ($\Delta \alpha$) experienced by the incoming light.[329] Reprinted with permission from Springer-Nature. (c) Fano-resonant gold metasurface for sub-100 fs all-optical switching using TPA. IRF stands for instrument response function.[331] Reprinted with permission from Wiley. (d) Ultrafast switching in dielectric GaP nanostructures designed as seen in the SEM image (left). The scale bar is 1 μm. Dynamics of the system were evaluated using pump-probe spectroscopy with sub-10 fs laser pulses, peak energy dependence of the differential reflectivity magnitude (middle) and the temporal dependence of the differential reflectivity (right).[332] Reprinted with permission from AAAS. (e) ITO nanowires pumped at its trans-LSPR in the near infrared (1500 nm wavelength) and probed in the visible range. SEM image (scale bar, 2 μm), electric field distribution (scale bar, 200 nm), normalized with regards to the incoming wave, and differential transmission spectrum at varying pump-probe delay time.[327,334] Reprinted with permission from Springer-Nature. (f) Graphene layer coupled to a gold waveguide for enhanced saturable absorption. Schematic, SEM image (scale bar, 2 μm), and all-optical switching extinction ratio as a function of pump pulse energy.[337] Reprinted with permission from Springer-Nature.

A combination of the last two approaches might pave the way to reach sub-fs and sub-fJ functionality at the same time. Furthermore, low-dimensional material systems such as quantum dots (QDs), graphene, and 2D semiconductors provide an unprecedented avenue to revolutionize information and communications technologies via recently discovered optical nonlinear properties.[352–354]

Nonlinear optics can be used for generating ultrashort pulses, laser spectrum conversion, ultrafast optical switching, and all-optical signal processing. By combining nonlinear optical processes in plasmonic systems with the dynamics of exciton-polaritons, it might be possible to achieve all-optical transistor functionalities with both PHz bandwidth and sub-fJ switching energy. To date, reaching these two goals at the same time has not yet been demonstrated, although switching mechanisms have been exploited in photonic systems made of inorganic semiconductor and displaying exciton-polaritons enabling transistor functionalities at cryogenic temperatures (see **Fig. 11a,b**).[335] Interestingly, by replacing inorganic semiconductors with an organic semiconductor in an optical microcavity, Zasedatelev *et. al* recently demonstrated that is possible to realize room-temperature operation of a polariton transistor through vibron-mediated stimulated polariton relaxation, with a sub-ps switching time, cascaded amplification and all-optical logic operation, however using micrometer-sized cavities (see **Fig. 11c-f**).[336] They also recently utilized stable excitons dressed with high-energy molecular vibrations to allow single-photon nonlinear operations at ambient conditions.[355] This pivotal result might open new horizons for practical implementations like sub-fs switching, amplification and all-optical logic at the fundamental quantum limit.

A key aspect of high-speed optical device application is the capability of ultrafast modulation and large modulation depth of high harmonic generation (HG)[356]. Many of these materials possess ultrafast photoexcitation dynamics and recovery time[357] to facilitate ultrafast control of nonlinear optical phenomena. However, their ultrathin nature makes it challenging to get reasonable efficiency in devices or to allow effective control of light at the nanoscale. This inability to compress light in the nanometric volume and a large momentum mismatch between the far-field photon and the excited optical mode, together with the ultra-thin material thicknesses motivate research efforts to combine these materials with plasmonic resonators and cavities to increase their light-matter interactions.[358,359] The use of plasmonic cavities and resonators together with low-dimensional materials leads to the generation of highly confined and strong EM fields, which can produce nonlinear optical responses such as harmonic generation (second and third harmonic generations, SHG and THG) at relatively low excitation powers (< 3 mW).[360–365] For example, Ren *et al.* reported a SHG enhancement factor of almost 10-fold from $AgInS_2$ QDs embedded in a Ag nanoparticles array under resonance excitation condition (when the LSPR of the Ag nanoparticles' array matches the excitation laser).[365] However, the sensitivity of HG may become even more prominent under harmonic resonance condition, when the LSPR of the plasmonic system overlaps well with the SHG wavelength of the nanomaterial due to the dipole (SHG) – quadrupole (plasmon) interaction.

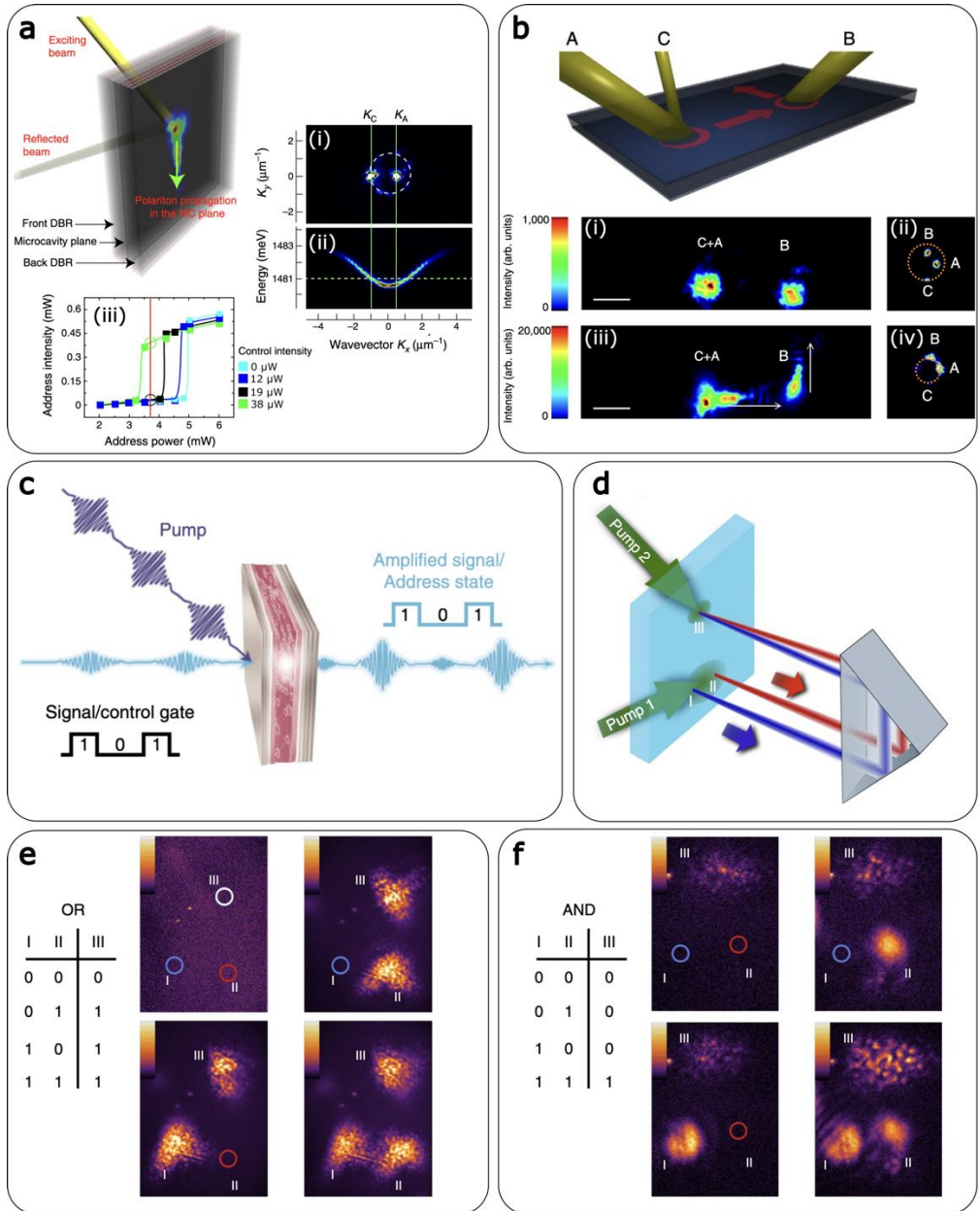

**Figure 11. All-optical transistor.** (a) Sketch of the experimental configuration in the work by Ballarini et al.[335] The exciting laser beam is directed with an angle on the microcavity sample and partially reflected from the front distributed Bragg reflector (DBR). Polaritons are created within the microcavity plane and propagate with a finite velocity. The one-to-one coupling with the external photons allows the observation of the polariton propagation through the emitted intensity from the back side of the sample. (i) Two-dimensional momentum space image of the emission under high excitation power shows the polariton states created by two laser beams impinging at different angles ($K_C$ and $K_A$) and same energy. (ii) Polariton dispersion along the direction corresponding to $K_y = 0$ of the upper panel. (iii) The transistor operation is briefly summarized. Address intensity plotted as a function of the Address power for different intensities of the Control: green line ($I_C = 38$ μW), black line ($I_C = 19$ μW), blue line ($I_C = 12$ μW) and without control (light blue line). The red line indicates the power of the Address for which a small change in the Control density brings the Address state from off (black circle) to on (green circle). (b) Illustration of the transistor setup, with two address beams A and B and a control C active at the same location as A. The polariton propagation is indicated with arrows. (i) Emission intensity colormap with both addresses as well as the control below the power

thresholds required for resonance. (ii) Far field emission intensity image in momentum space. (iii) Similar image as (i) with C slightly increased. The emission intensity here about 20 times larger than in (i). (iv) Momentum space image, showing A and B with enhanced emission intensities whilst C is negligible.[335] Reprinted with permission from Springer-Nature. (c) Schematic of an all-optical organic polariton transistor. (d) Illustration of AND and OR gate operations, with two control pulses regulating addresses I and II. The pump at III is tweaked such that one (two) amplified addresses is necessary to excite the structure at III. (e) Real-space photoluminescence of the transistor with the OR setup and (f) with the AND setup. Working operations are demonstrated in both cases.[336] Reprinted with permission from Springer-Nature.

As demonstrated by Han *et al.*, overlapping of the LSPR of Ag nanocubes with the propagating SP on Ag film/$Al_2O_3$ system (nanoparticles on mirror system) with the SHG wavelength at 410 nm from $WS_2$ monolayer resulted in up to 300-fold enhancement of SHG signal at an excitation power below 100 µW (see **Fig. 12a**).[364] Moreover, a 7000-fold enhancement in the SHG response has been observed when using a plasmonic platform coupled to a $WS_2$ monolayer.[366] At resonant coupling strength (when the LSPR couples resonantly with the exciton, i.e., in the strong coupling regime), photons confined in the nanometric volume coherently exchange energy with excitons at a rate higher than that of the dephasing processes (i.e. the rate at which photons escape from the cavity or exciton de-phases) creating a nonlinear hybrid quantum state, called polaritons. Such polaritonic states thus, open new avenues like Bose-Einstein condensation,[367] polaritonic lasing,[368] optical parametric amplification,[369] Rydberg excitonic phenomena,[370] etc. in TMDs to be investigated at near room temperatures owing to the large exciton binding energies in them. Investigation of ultrafast dynamics of these excitonic and polaritonic processes in 2D semiconductor/plasmonic heterostructures is a growing research area with an opportunity to provide a deeper or new understanding of energy exchange effects between plasmons and excitons on the ultrafast timescale under deep-subwavelength optical confinement.[338,371,372] For example, Tang *et al.*[373] studied the ultrafast dynamics of the nonlinear response from $WS_2$ monolayer coupled to silver nanodisks in the strong coupling regime. They observed that with increasing fluences, plasmon-exciton polaritons (plexcitons) become weaker. In **Fig. 12b** it is shown a series of reflection spectra demonstrating coherent plexcitons formation at 100 fs under various pump incident fluences. The upper and lower branch modes of the plexcitons systematically shift towards the exciton resonance with increasing fluence resulting in reduction of spectral splitting, and exhibit considerable linewidth broadening, bringing about an apparent rise of the splitting window (increasing valley). Their pump-probe measurement unveiled a giant plasmon induced nonlinearity with non-equilibrium exciton-plasmon processes over distinct time scales governed by 1) repulsive Coulomb interaction between excitons and 2) Pauli blocking resulting in excitonic saturation and reducing exciton-photon coupling strength[359].

Besides their outstanding capability of light-trapping and concentrating/amplifying an EM field in nanometric volumes, plasmonic nanostructures can also convert light into electrical energy by generating hot electrons.

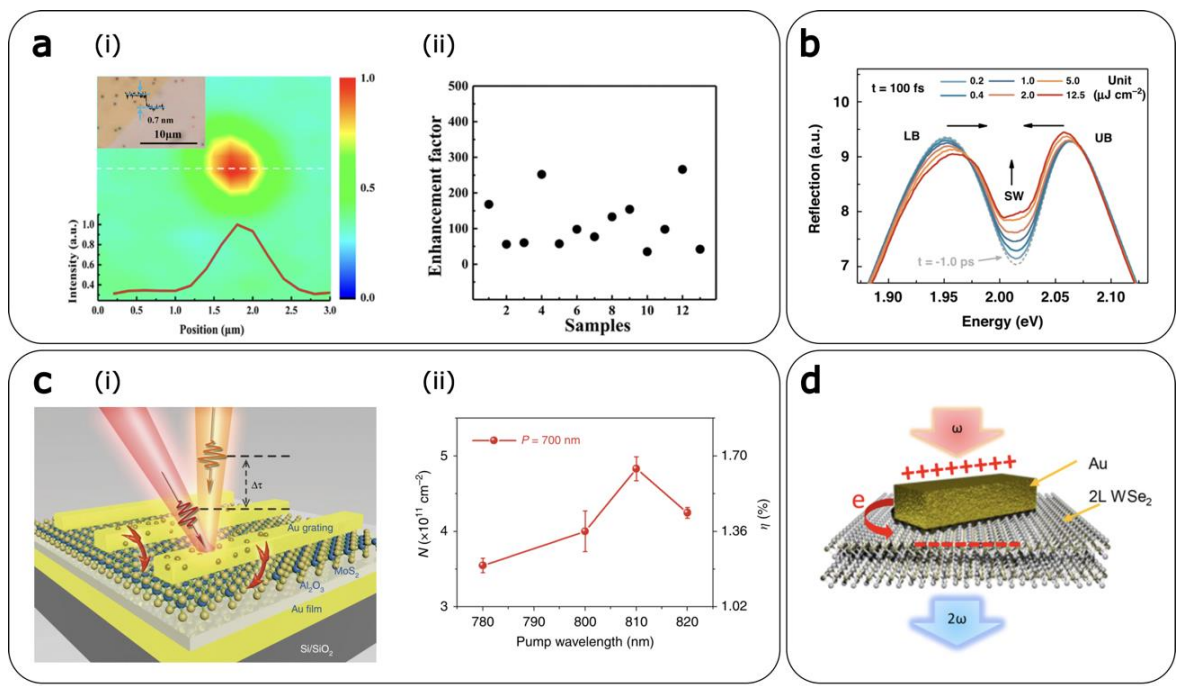

**Figure 12. Application of ultrafast plasmonics in low-dimensional semiconductors.** (a) (i) Normalized SHG intensity map of a monolayer $WS_2$ integrated in a plasmonic nanocavity with a scanning step of 200 nm. The curve is the SHG intensity line profile along the white dashed line. Inset shows an optical micrograph of the hybrid system with black dots are the Ag nanocubes. (ii) SHG enhancement factor with different Ag nanocubes under fixed excitation laser with wavelength of 820 nm.[364] Reprinted with permission from ACS. (b) Transient reflection spectra of the $WS_2$/Au-nanodiscs plexciton system measured at 100 fs at different fluence rate. The grey dashed line is the unaffected probe spectrum by probe pulse at -1 ps.[373] Reprinted with permission from Springer-Nature. (c) (i) Schematic of pump-probe measurements on Au grating/$MoS_2$/$Al_2O_3$/Au sandwiched heterostructure to demonstrate plasmonic hot electron transfer from a gold grating to the MoS2 monolayer. (ii) Pump wavelength dependent injected hot electron densities using the previous setup from (d) and corresponding efficiency for the heterostructure with 700 nm grating period.[374] Reprinted with permission from Springer-Nature. (d) A diagrammatic model to describe the hot electron induced symmetry breaking and probing SHG in a Au/2L $WSe_2$ hybrid system.[375] Reprinted with permission from ACS.

As mentioned in **Section 3**, upon optical absorption and plasmon excitation in the nanostructures, the decaying plasmon field transfers accumulated energy (via Landau damping) to the conduction band electrons. The efficiency of hot electron generation depends on multiple factors[376]. Among them, the most important one is the LSPR effect. At the LSPR resonance frequency the photon excitation produces the maximum number of highly energetic electrons ready to escape the metal surfaces. Thus, hot electrons can transfer into the unoccupied states of nearby semiconductors, when put in contact with the metal nanostructures that are being resonantly excited at the LSPR frequency[376,377] (see also **Section 3**). For example, Shan et al.[374] demonstrated direct evidence of plasmonic hot electron transfer from a gold grating to a $MoS_2$ monolayer, that occurs at ~40 fs with a maximum efficiency of 1.65 %. The coupling between LSP of Au grating on top and SPP of Au film underneath (see **Fig. 12c**) facilitated coherent energy exchange and created a strong out-of-plane electric field, which decreases the radiative damping rate of LSPR and accelerates exportation of hot electrons into $MoS_2$. The net energy produced in the coupling process (the SP energy otherwise produces heat) is recycled to generate hot electrons via LSPR of the gratings. As a result, pumping of hot electrons and the consequent injection into the $MoS_2$ monolayer was amplified as shown in **Fig. 12c**.

Harvesting plasmonic hot electrons into a semiconductor has multi-faceted applications. They can be used for light harvesting and optoelectronics,[378,379] probing mobility via Fröhlich interaction,[380] and photocatalysis as well as $H_2$ evolution.[381,382] Due to strong spatial confinement of 2D semiconductors, injection of hot electrons may introduce additional effects, which are unseen in bulk systems. For example, a giant bandgap renormalization in $WS_2$ (550 meV) at room temperature,[383] semiconducting to metallic phase transition in monolayer $MoS_2$[384,385] has been reported in the literature. Plasmonic hot electrons can also be used to break the inversion symmetry in 2D semiconductors leading to the observation of SHG signal. Controlling SHG is fundamental to nonlinear optical applications, which requires a non-zero 2nd order susceptibility tensor only observed in a non-centrosymmetric system. Therefore, generating SHG signal in a centrosymmetric bilayer TMDs systems was endeavored via breaking the inversion symmetry by an external electric field.[386] However, this method of controlling SHG cannot occur at ultrafast timescale, since required voltage is typically high and at high operating frequencies current becomes unsustainable. Wen et al.[375] reported the symmetry breaking in bilayer $WSe_2$ using plasmonic hot electron injection, which works on a similar principle but at an ultrafast timescale (principle of SHG from bilayer WSe2 is schematically presented in **Fig. 12d**). The injected electrons create an asymmetric electric field inside the bilayer, causing a symmetry breaking and induce SHG signal, which is also aided by the presence of LSPR induced strong local field at the proximity. The recorded rise and fall time of hot electrons were 119 fs and 1.84 ps, respectively, indicating great promise for ultrafast nonlinear applications.

## 6. Ultrafast plasmonics in the relativistic regime

In the former sections, light pulses with low intensity were utilized to trigger reproducible plasmon-driven physical and chemical processes in various types of nanostructures. When laser intensity reaches $10^{11}$ - $10^{13}$ W/cm$^2$, irradiation of nano-objects in vacuum leads to the emission of low energy (1-100 eV) electrons.[387,388] Under certain conditions, these electrons can interact with a resonant state of the nanostructures themselves, i.e., either LSPs[389] or SPPs,[390] which can further accelerate them up to 1 keV. Besides providing insight into the fundamental processes that take place in this regime,[8,391] these electrons are well suited for various applications, such as U-TEM, as already discussed in **Section 2**.[392]

By increasing the laser intensity further, the material starts to be significantly ionized and optically damaged. This way sub-optical cycle metallization of $SiO_2$ nanospheres has been demonstrated by measuring an increased cutoff energy of the emitted electrons in few-cycle laser pulses around $2\times10^{14}$ W/cm$^2$ intensity.[393] Already at these low ionization levels, the temporal change of electron density plays an important role as it strongly limits the resonance in time (when the plasma frequency is equal to the laser frequency) and sets the field enhancement factor to about 3 when the critical electron density is surpassed. If the laser intensity is further increased beyond the optical damage threshold, the plasma, that optimally is still nm sized, dominates the interaction with the laser pulse. This is a hot, overdense ($n_e \gg n_{crit}$, where $n_{crit}$ is the critical density) and spatially strongly inhomogeneous plasma, which quickly expands in space and reaches µm extension within a few ps after its generation. A fundamental laser property in the ultrahigh laser intensity regime determining the generation time of the plasma is the high-dynamic-range temporal intensity contrast.[394] It describes the laser intensity at a certain time instant normalized to the peak temporal intensity of the pulse. In practice, depending on the details, $10^6$-$10^{10}$ times lower intensity than the peak value can ionize the target and generate plasma. For typical lasers, these levels are reached not only ps, but even ns before the fs laser pulse and thus the limited contrast of modern high-intensity lasers strongly limits their application for nanophotonics and therefore also the number of published papers. Various methods such as the generation of the second harmonic and the utilization of the so-called plasma mirror can improve the contrast of lasers at the cost of lower pulse energy. Another alternative is light sources based on Optical Parametric Chirped-Pulse Amplifiers (OPCPA), which typically have excellent contrast.[395]

Atomic clusters, an intermediate form of matter between molecules and solids, with typical diameters of a few nm, have been used as target to investigate laser-cluster interactions. Clusters are generated by a gas jet plume, originating from a cryogenically cooled gas nozzle, expanding into vacuum and cooling during expansion. The interaction of moderately high intensity laser pulses ($10^{16}$ – $10^{17}$ W/cm$^2$) with gas-phase cluster targets have been investigated under different conditions. Coulomb explosion of hydrogen clusters has yielded protons with 8 keV kinetic energy in good agreement with expectations.[396] However, xenon cluster explosions have produced xenon ions with kinetic energies up to 1 MeV, much beyond ion energies from molecular or small cluster targets.[397] This is explained by the fact that the expansion of the clusters reduces the electron density to a point where the plasma frequency is resonant with the laser frequency. By applying deuterium clusters Ditmire *et al.* have observed D-D fusion from Coulomb exploding clusters and created large number of fusion neutrons ($10^5$ neutrons/J laser energy) with a table top laser.[398]

At an intensity of about $10^{18}$ W/cm$^2$ (at ~1 µm wavelength) a free electron starts to oscillate in the laser field with relativistic velocity, therefore this value is the *relativistic intensity* limit. Due to the demanding temporal contrast requirements, only few works have been performed at such a high intensity with nanoobjects. One of the existing lines of investigation studied the interaction between the high fields of the laser and propagating surface plasmons. To this end, optical gratings were utilized as targets representing nm extension in one dimension and nm modulation in another one, as schematically shown in **Fig. 13a**. Certain high-order harmonics are predicted to be generated at the wavelength of the grating period and its harmonics by attosecond electron bunches parallel to the target surface in the case of perpendicular laser incidence.[399] It has been confirmed experimentally by Cerchez *et al.*,[400] by showing that high-order harmonics with parameter-dependent spectral composition from grating targets can be emitted under grazing emission angles separated from the laser fundamental radiation. Grating targets irradiated by $5 \times 10^{19}$ W/cm$^2$ laser intensity under resonant conditions resulted in the generation of 100 pC electron bunches with 5-15 MeV energy along the target surface by propagating relativistic surface plasma waves.[401] Flat targets under the same conditions produced much less, lower energy electrons as shown in **Fig. 13b**. Resonant excitation of surface waves on thin grating targets also increases the laser absorption and proton energies by a factor of 2.5 from the target normal sheet acceleration as demonstrated by Ceccotti *et al.*.[402]

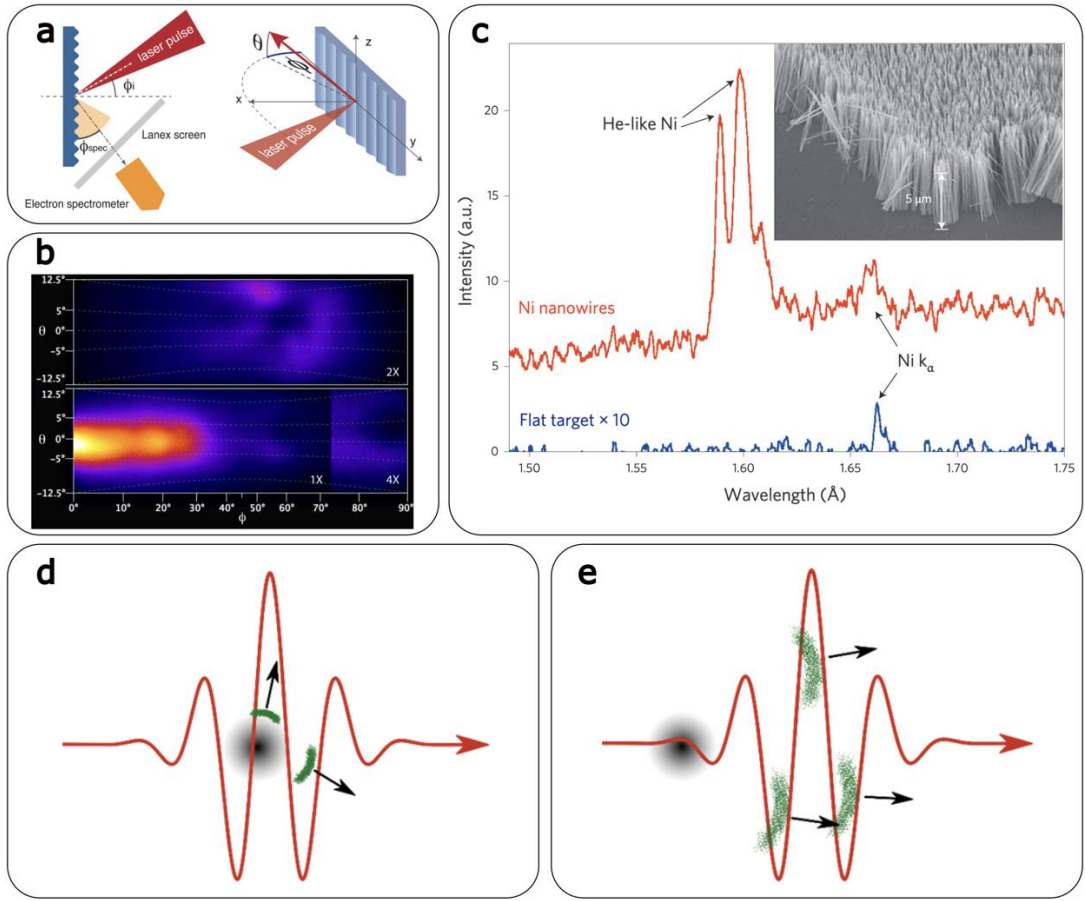

**Figure 13. Different types of relativistic surface plasma experiments.** (**a**) Setup for generating propagating surface plasmons along a grating. (b) Angularly resolved electron emission measured when using a flat surface (top) and a grating (bottom) as targets. A filter placed in front of the detector cut off electrons with energies below 1.3 MeV. The upper and the right bottom panels have been amplified 2x and 4x compared with the bottom left to highlight features in the signals.[401] Reprinted with permission from APS. (c) Measured single-shot X-ray emission when utilizing an array of nanowires (red), shown in the inset, and a flat surface (blue) for comparison. Notice that besides the increase in signal corresponding to the $K_\alpha$ line, the array of nanowires shows a strong He-like Ni signature, highlighting the high degree of ionization and plasma density.[403] Reprinted with permission from Springer-Nature. Figures (d, e) show the first and second step of relativistic electron acceleration from nanodroplets. First, (d) an electron bunch is extracted and accelerated away from nanodroplet in a direction determined by Mie scattering for every half cycle of the driving laser. Then, (e) the bunches are further accelerated in the laser propagation direction by vacuum-laser acceleration.

Another line of experiments focused in the generation of volumetric plasmas that are simultaneously dense and hot. In general, a relativistic laser pulse interacting with solid matter creates a layer of plasma at the surface of the material, which stays overdense for the typical time duration of the pulse (5 – 100 fs). Once established early in the interaction, this overdense plasma prevents the remaining of the incoming laser pulse to penetrate and heat it up deeper than typically a few 100 nm. Effectively, this limits the volume of matter that is efficiently heated and challenges the creation of extreme laser-generated plasmas. However, it has been shown that this constraint can be cleverly avoided when an array of vertically aligned nanowires, shown in **Fig. 13c**, is used as laser target.[403] The incident laser pulse propagates along the space between the nanowires, which are about 10 µm long, producing higher electron density ($100n_{crit}$), higher temperatures (multi-keV), higher laser absorption, and increased x-ray radiation than when utilizing flat targets under the same conditions. This technique has been utilized to accelerate ions and realize deuterium-deuterium fusion with increased neutron yield.[404,405]

The last type of investigations is based on single 2D and 3D nanotargets. Numerical simulations have predicted attosecond electron bunches with relativistic energy from the interaction of (overdense) plasma nanodroplets with relativistic intensity laser pulses.[406,407] The interaction is described by a two-step model as illustrated in Fig. **Fig. 13d, e**.[408,409] The first step is the nanophotonic emission of relativistic electrons from the target by the localized surface plasmon and the enhanced local fields. This step is mainly governed by the Mie theory and the emission directions correspond to the angles of maximum field enhancement. The second step is vacuum-laser acceleration (VLA) of these electrons that co-propagate with the laser pulse. Electrons propagate in two main directions on the two sides near the laser direction. This final angular distribution of the electrons is mainly determined by VLA, although the first step has also some influence. Utilizing few-cycle laser pulses, a relativistic isolated attosecond electron bunch is generated on one side, though the process is carrier-envelope phase (CEP) dependent.[410] These nanodroplets, after the electrons are removed by the laser, have been suggested for Coulomb-explosion-dominated ion acceleration to (laser intensity-dependent) few MeV – few10 MeV ion energy using two-cycle laser pulses.[411]

In an experiment (**Fig. 14a**), nanotips with 100 nm tip diameter have been irradiated with $6\times10^{19}$ W/cm$^2$ intensity sub-two-cycle pulses and measured intensity-dependent electron propagation angle, corresponding to the relativistic ponderomotive scattering angle.[409] This supports the VLA process, though, slight direction change was also detected when the target size was changed. The electron energy spectrum was broad, peaked at 3-5 MeV and extended up to 9 MeV (**Fig. 14b**), indicating relativistic electron bunches with charges up to 100 pC (~$10^9$ electrons) injected and accelerated within a half optical cycle (in fair agreement with the simulations shown in **Fig. 14c**). CEP dependence of the propagation angle as well as angular distribution have been demonstrated as shown in **Fig. 14d**. Strong indication of >TV/m fields accelerating the electrons has been presented that significantly surpasses former values.

## 7. Conclusions and future perspectives

In this review we have explored the current state and future prospects of the field of ultrafast plasmonics, ranging from fundamental science to technological aspects. We have focused the reader attention to the state-of-the-art of modelling methods and experimental techniques that have been developed over the last decade to unveil the interaction of ultrafast pulses with nanostructured optical materials made of noble metals, highly doped semiconductors, and hybrid plasmonic-dielectric/low-dimensional semiconductors and/or molecular systems. To this end, we have showcased most recent works to uncover the light-matter interaction with the highest possible spatial and temporal resolutions. The ability to measure and understand the ultrafast response of materials on the nanoscale is crucial for future development of functioning devices. Therein, pushing the time-resolution of experiments by employment of ever shorter probe pulses, even below one femtosecond, will allow to directly study fundamental material responses on the level of the cycles of light. Further, it is now possible for experimentalists to correlate such fundamental processes, like absorption or emission of light, at the single-particle level, promising many exciting discoveries in ultrafast plasmonics related to the quantum nature underlying the excitation of the material.[412]

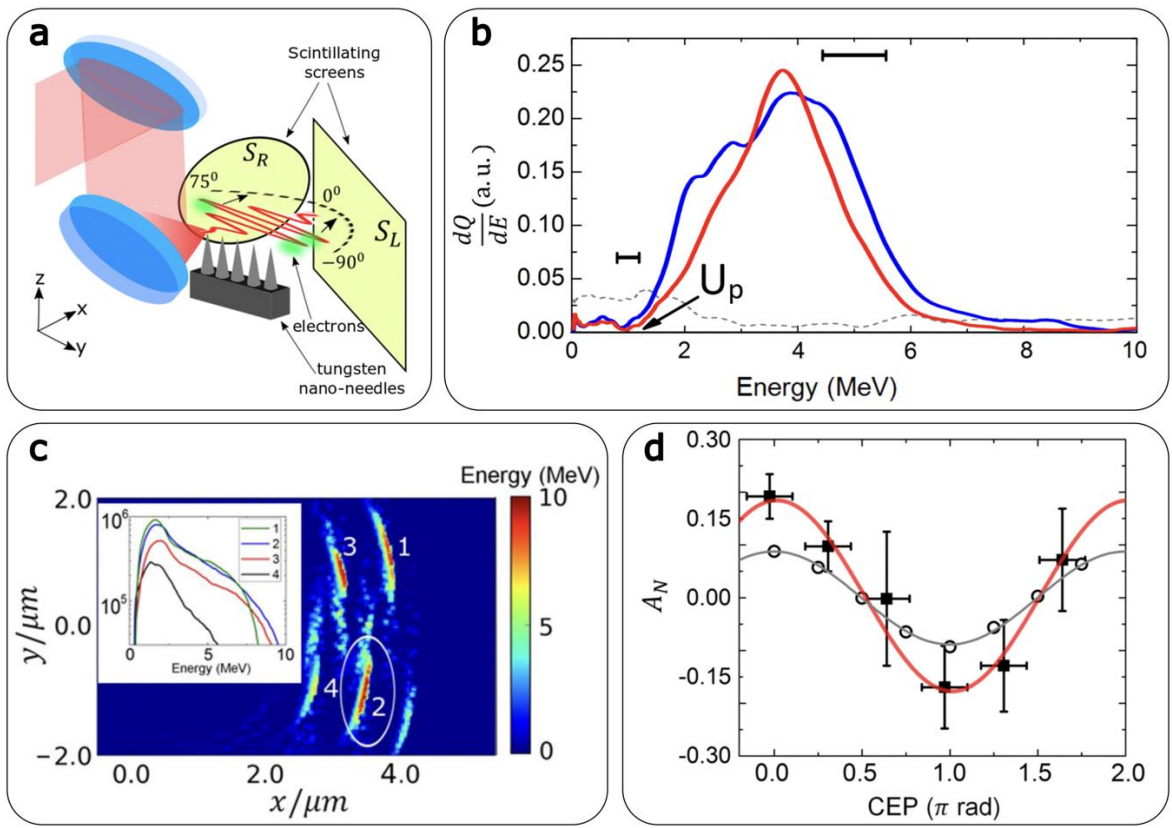

**Figure 14. Acceleration of electrons from isolated nanotips in the relativistic nanophotonics regime.** (a) Setup for relativistic electron acceleration using nanotips. (b) Experimental energy spectrum of the accelerated electrons. (c) Particle in a cell simulation of the emitted electron bunches, with the predicted energy spectra of every bunch as an inset. (d) Correlation between asymmetry in the electron emission direction and CEP of the driving laser.[409] Reprinted with permission from Springer-Nature.

In addition, because of the different length and time scales involved in the plasmon relaxation dynamics, we have pointed out that only a proper synergistic cooperation between state-of-the-art experimental and theoretical methods can be effective for achieving an understanding of such complicated dynamical processes. From a theoretical standpoint, the current scientific interest is directed towards extending real-time *ab-initio* methods in combination with multiscale approaches, so to make it computationally feasible the analysis of phenomena that span different length and time scales, such as hot carriers transport properties of plasmonic materials, thus not only focusing on plasmonic absorption and subsequent generation of hot carriers.[170,184] This aspect, which cannot be adequately treated without an appropriate inclusion of the real material atomistic structure,[169] is crucial for a rational design of plasmonic nanomaterials with applications in photocatalysis, energy storage, and so forth. We have also showcased how plasmonic nanostructures have an unprecedented potential to convert light into chemical energy, addressing thus fundamental issues of our modern society, such as decreasing the atmospheric greenhouse gasses and developing new technologies for sustainable energy production. To this end, understanding the ultrafast dynamics of non-thermal (or hot) charge carriers as well as the electron-phonon and phonon-phonon interactions is crucial for developing new efficient plasmonic catalysts. Although the hot carrier dynamic in purely plasmonic systems has been unraveled by ultrafast spectroscopy, this is not yet the case for more complex (and relevant) interfaces, such as hybrid plasmonic catalysts (i.e., metal-metal, metal-semiconductor, or metal-perovskite systems).[200] Thus, exploring the phonon dynamics and hot carrier relaxation mechanisms in plasmonic hybrid systems will help decode the roles of temperature and non-thermal charge carriers for driving and enhancing chemical reactions. This new direction of research can lead to completely new plasmonic hybrid materials in the upcoming years with exciting properties for energy conversion.[413] Furthermore, we have shown how the combination of nanophotonics, magnetoplasmonics, and spintronics opens new possibilities for the practical implementation of magnetic field-controllable nanoscale devices for ultrafast data processing and storage. In this context, plasmonics might enable all-optical magnetization switching to reach the atomic scale.[414–419] By exploiting the non-thermal dynamics of the electron gas, we expect that novel plasmon-driven phenomena can be discovered and controlled for the duration of the laser pulse (sub-10 fs), where the microscopic degrees of freedom, such as the spin, can be strongly coupled to the amplitude, frequency, and polarization of the plasmonic field. This direction is still unexplored and can become a rising research line in the upcoming years, as it might unveil novel pathways to control spin dynamics in the non-thermal regime also with metallic nanostructures, thus overcoming the intrinsic limitations placed by ohmic and other losses, opening excellent opportunities towards plasmon-driven ultrafast magneto-optics. We have also shown that plasmonics can serve as an important tool to push towards low-power all-optical ultrafast processing applications. Pure plasmonic platforms already offer the possibility of sub-100 fs all-optical modulators, however at the cost of relatively high control energies and offering only low switching contrasts. Recent investigations that combine plasmonics with other nonlinear elements, such as graphene, ITO, or ultrathin semiconductor materials with strong excitonic effects, suggest a promising path toward the development of efficient ultrafast all-optical switches. A route that has not yet been addressed for this purpose is the formation of hybrid systems incorporating high-index dielectric nanoscale resonators, which could help bringing down the required control powers due to their low losses, high optical nonlinearities, and field confinement abilities. We have also summarized some representative nonlinear optical applications of low-dimensional semiconductors incorporated with plasmonic materials, which offer great promise for novel ultrafast optical applications.[420,11,421] Although their ultrathin thickness and large momentum mismatch with photon provide a major challenge to tame light into nanometric volume, we predict new breakthrough experiments combining these materials with plasmonic and dielectric nanophotonic systems in coming years, thus driving the field into new directions. In particular, special emphasis should be given to controlling the optical damage threshold, influence of intrinsic (such as doping, crystal structure, and vdW supercell) as well as extrinsic (such as electric field, strain) parameters on nonlinear optical properties in this novel class of materials, and finally, care must also be taken to developing novel device fabrication processes that preserve the materials' properties. Finally, we gave a snapshot of ultrafast plasmonics in the relativistic realm. The latter is a promising field of science with potential applications in fusion research, or as novel particle and light sources providing properties such as attosecond duration that is not available from other sources nowadays.

**Acknowledgements.** N.M. acknowledges support from the Swedish Research Council (grant n. 2021-05784) and Kempestiftelserna (grant n. JCK-3122). N.M. and S.C. acknowledge support from the European Innovation Council (grant n. 101046920 'iSenseDNA'). N.M. and D.B. acknowledge support from the Luxembourg National Research Fund (grant n. C19/MS/13624497 'ULTRON'). N.M., D.B. and S.C. acknowledge support from the European Commission (grant n. 964363 'ProID'). D.B. acknowledges support from the European Research Council (grant n. 819871 'UpTempo') and the European Regional Development Fund Program (grant n. 2017-03-022-19 'Lux-Ultra-Fast'). L.V. acknowledges support from the Swedish Research Council (grants n. 2019-02376 and n. 2020-05111), Knut och Alice Wallenberg Stiftelse (grant n. 2019.0140), and Kempestiftelserna (grant n. SMK21-0017). A.S. and E.C. acknowledge German Research Foundation under e-conversion Germany´s Excellence Strategy (grant n. EXC 2089/1 – 390776260), the Bavarian program Solar Energies Go Hybrid (SolTech), the Center for NanoScience (CeNS) and the European Research Council (grant n. 802989 'CATALIGHT'). G.G. acknowledges funding from Agencia Nacional de Promoción de la Investigación, el Desarrollo Tecnológico y la Innovación (grant n. PICT 2019-01886), Consejo Nacional de Investigaciones Científicas y Técnicas (grant n. PIP 112 202001 01465) and Universidad de Buenos Aires (project n. 20020190200296BA). D.J. acknowledges partial support by the Asian Office of Aerospace Research and Development of the Air Force Office of Scientific Research (grants n. FA2386-20-1-4074 and n. FA2386-21-1-4063). M.R. acknowledges support from the German Research Foundation for Walter Benjamin Fellowship (award n. RA 3646/1-1). D.G. acknowledges support from the European Commission (grants n. 964995 'DNA-FAIRYLIGHTS').